\documentclass[aps,prb,reprint,
showpacs,superscriptaddress,floatfix]{revtex4-2}

\usepackage{amsmath,amssymb,amsfonts}
\usepackage{mathtools}
\usepackage{graphicx}
\usepackage[colorlinks=True,linkcolor=red,citecolor=blue,urlcolor=blue]{hyperref}
\usepackage{bookmark}
\usepackage[dvipsnames]{xcolor}
\usepackage{braket}
\usepackage{nicefrac}
\usepackage{bm}
\usepackage{bbm}
\usepackage{braket}
\usepackage{dcolumn}
\usepackage{slashed}
\usepackage[normalem]{ulem}
\usepackage{array}
\usepackage{makecell}
\usepackage{subfigure}
\newcolumntype{C}{>{$}c<{$}}

\def\infinity{\infty}

\begin{abstract}
We investigate domain-wall physics in a non-Hermitian Su-Schrieffer-Heeger model featuring the non-Hermitian skin effect and its higher-dimensional extensions, focusing on analytical eigenstate solutions under open boundary conditions.
By gluing two Su-Schrieffer-Heeger chains with inverted coupling ratios together, we identify two distinct interface geometries, and derive closed-form quantization conditions for the complex wave number and wave functions using symmetry-based ansatzes.
Besides conventional skin modes and topological zero modes, the domain walls generate additional localized states at finite energy that detach from the bulk open-boundary-condition continuum.
We validate the analytical results by exact diagonalization, and further generalize the interface construction to a two-dimensional Lieb lattice, where competing nonreciprocities produce tunable funneling regimes towards codimension-one and codimension-two interfaces.
\end{abstract}

\begin{document}
\title{Exact solutions of nonreciprocal Su-Schrieffer-Heeger model with domain walls}
\author{Tong Wang}
\email{tong.wang@mpl.mpg.de}
\thanks{These authors contributed equally to this work}
\affiliation{Max Planck Institute for the Science of Light, 91058 Erlangen, Germany}
\author{Mengjie Yang}
\email{mengjie.yang@u.nus.edu}
\thanks{These authors contributed equally to this work}
\affiliation{Department of Physics, National University of Singapore, Singapore 117551, Singapore}
\author{Flore K. Kunst}
\email{flore.kunst@mpl.mpg.de}
\affiliation{Max Planck Institute for the Science of Light, 91058 Erlangen, Germany}
\affiliation{Department of Physics, Friedrich-Alexander-Universit\"at Erlangen-N\"urnberg, 91058 Erlangen, Germany}
\date{\today}

\maketitle

\section{Introduction}
A hallmark of topological phases of matter is the presence of boundary-localized eigenstates, which originate from nontrivial bulk topology~\cite{kane2005quantum,kane2005z,fu2007topological,hasan2010colloquium,moore2010birth,ryu2010topological,qi2011topological,ando2013topological,bernevig2013topological,khanikaev2013photonic,chiu2016classification,asboth2016short}. Recent work on non-Hermitian~(NH) systems reveals the additional localization of a macroscopic number of eigenstates on the boundaries, known as the NH skin effect, which points toward the breakdown of the conventional bulk-boundary correspondence (BBC)~\cite{lee2016anomalous,alvarez2018non,kunst2018a,yao2018edge,kunst2019a,lee2019anatomy,kunst2019non,lee2019hybrid,yokomizo2019non,longhi2019probing,song2019non,lin2023topological,wang2024non,yang2025non}. This effect is an inherent phenomenon in NH systems without counterpart in Hermitian systems~\cite{li2020critical,ashida2020non,okuma2020topological,bergholtz2021exceptional,longhi2019metal,mandal2020nonreciprocal,zhang2020non,hofmann2020reciprocal,longhi2019topological,ghatak2020observation,zhang2021observation,weidemann2022topological,xiao2020non,yang2024percolation,lin2024observation,li2024observation,xiao2024observation, zhang2020correspondence, Gohsrich2025,yang2026beyond}, and a variety of novel phenomena have been reported in connection with the NH skin effect, such as the modified BBC through introducing the biorthogonal polarization~\cite{kunst2018a,kunst2019non}, generalized Brillouin zone~\cite{yao2018edge,yokomizo2019non,zhang2020correspondence}, or doubled Green’s function~\cite{borgnia2020non}, 
the unconventional localization transition from the interplay of quasiperiodicity and NH skin effect \cite{longhi2019metal,longhi2019topological,weidemann2022topological}, nonreciprocal and directional transport in driven-dissipative systems~ \cite{mcdonald2018phase,mandal2020nonreciprocal,okuma2021quantum, liang2022dynamic,yang2026reversing,yang2025beyond,Rao2026Transport,yang2026nonlocality}, and extremely sensitive sensors~\cite{Budich2020Non,Mcdonald2020sensing,Koch2022,Yuan2023,Konye2024Non,Deng2024sensor,Parto2025enhanced}.

\begin{figure}[t]
\centering
\includegraphics[width=0.9\linewidth]{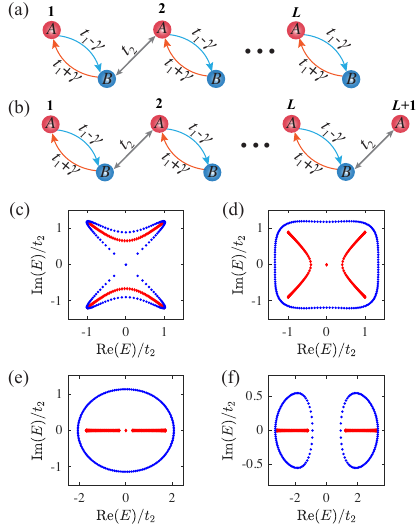}
\caption{The non-Hermitian SSH model with (a) unbroken unit cells and (b) a broken unit cell on the right end. The PBC (blue dots) and OBC spectra (red diamonds) for a lattice of $80$ unbroken unit cells are shown in (c)-(f). The model parameters are $\gamma/t_2=1.2$ and (a) $t_1/t_2=0.2$, (b) $t_1/t_2=0.8$, (c) $t_1/t_2=1.4$, (d) $t_1/t_2=2.5$.}
\label{fig_pbc_obc}
\end{figure}

The NH skin effect can also induce a funneling effect, where the majority of bulk states accumulate at an interface between NH systems. This effect has been observed in a one-dimensional discrete-time quantum walk platform~\cite{weidemann2020topological} and in diffusive thermal systems~\cite{cao2021diffusive}. The funneling effect has a wide range of applications in optics and condensed matter physics, including light- and quantum-state-harvesting platforms~\cite{yang2022concentrated}, wave manipulation~\cite{parto2021non,Zhong2023plate}, robust directional transport~\cite{longhi2015non,longhi2015robust}, precision measurements~\cite{Schneider2025} and information processing~\cite{song2020two,lin2021steering}. Despite the progress in experiments, analytical studies of systems exhibiting the NH skin effect as well as the funneling effect remain scarce. Related interface problems can be treated using transfer-matrix approaches, where Fermi arcs on two-dimensional interfaces between stacked Weyl semimetals have been studied in the Hermitian limit~\cite{Dwivedi2018}, and in the context of non-Hermitian defect-state formalisms as done for a one-dimensional Su-Schreefer-Heeger (SSH) chain with staggered complex potentials in Ref.~\cite{schomerus2013topologically}. More recently, non-Bloch band theory has been extended explicitly to non-Hermitian domain-wall systems, where interface localization can be related to differences in spectral winding between adjacent domains~\cite{Xu2026DomainWall}. These studies focus on codimension-one interfaces, where the codimension refers to the difference between the dimension of the system and the dimension of the interface. In contrast, higher-dimensional systems with lower-dimensional interfaces (including interfaces with codimensions two and three) remain much less explored.

In this paper, we provide a thorough investigation of the funneling effect in systems described by an SSH model with asymmetric intracell coupling, often referred to as the nonreciprocal SSH chain, cf. Fig.~\ref{fig_pbc_obc}(a, b). We identify two types of domain-wall interfaces formed by gluing subsystems with inverted coupling ratios together, cf. Fig.~\ref{fig_domain_wall}, and derive exact analytical solutions for the full set of eigenenergies and eigenstates in the inversion-symmetric configuration, where the number of sites in each subsystem is equal. These interfaces give rise to additional zero-energy modes as well as finite-energy interface states detached from the bulk continuum under open boundary conditions~(OBCs). We connect the existence of these additional states to the symmetry and topological characterization of the nonreciprocal SSH chain~\cite{yao2018edge,okuma2020topological,kawabata2019prx}. We further present spectra and right-eigenstate wave-function profiles to demonstrate how the funneling effect depends on model parameters.

Going beyond the one-dimensional case, we also study a two-dimensional setup consisting of superimposed nonreciprocal SSH chains forming a two-dimensional Lieb lattice, cf. Fig.~\ref{fig:domain_wall_2D}. Depending on the choice of asymmetric hoppings, we create domains of codimension one and/or two, and we find a tunable funneling regime towards these interfaces. 

This paper is structured as follows: In Sec. \ref{sec2} we give a give a short overview of the nonreciprocal SSH model studied in this paper, highlighting the exact solutions of eigenstates in the broken- and unbroken-unit-cell case. In Sec. \ref{sec3} we introduce systems with different interfaces, type A and type B, by gluing two identical chains with inverted coupling ratios together. We focus on the case where the interfaces are located exactly at the center, and separate the eigenstates into two ansatzes. This enables us to derive the exact solutions of eigenstates with the help of results from Sec. \ref{sec2}. In Sec. \ref{sec4} we show numerical results of systems in both one- and two-dimensions with different types of interfaces and model parameters.  Finally, we summarize our results and discuss possible future directions.

\section{Nonreciprocal SSH model with open boundaries}
\label{sec2}

\begin{figure*}
\centering
\includegraphics[width=0.85\linewidth]{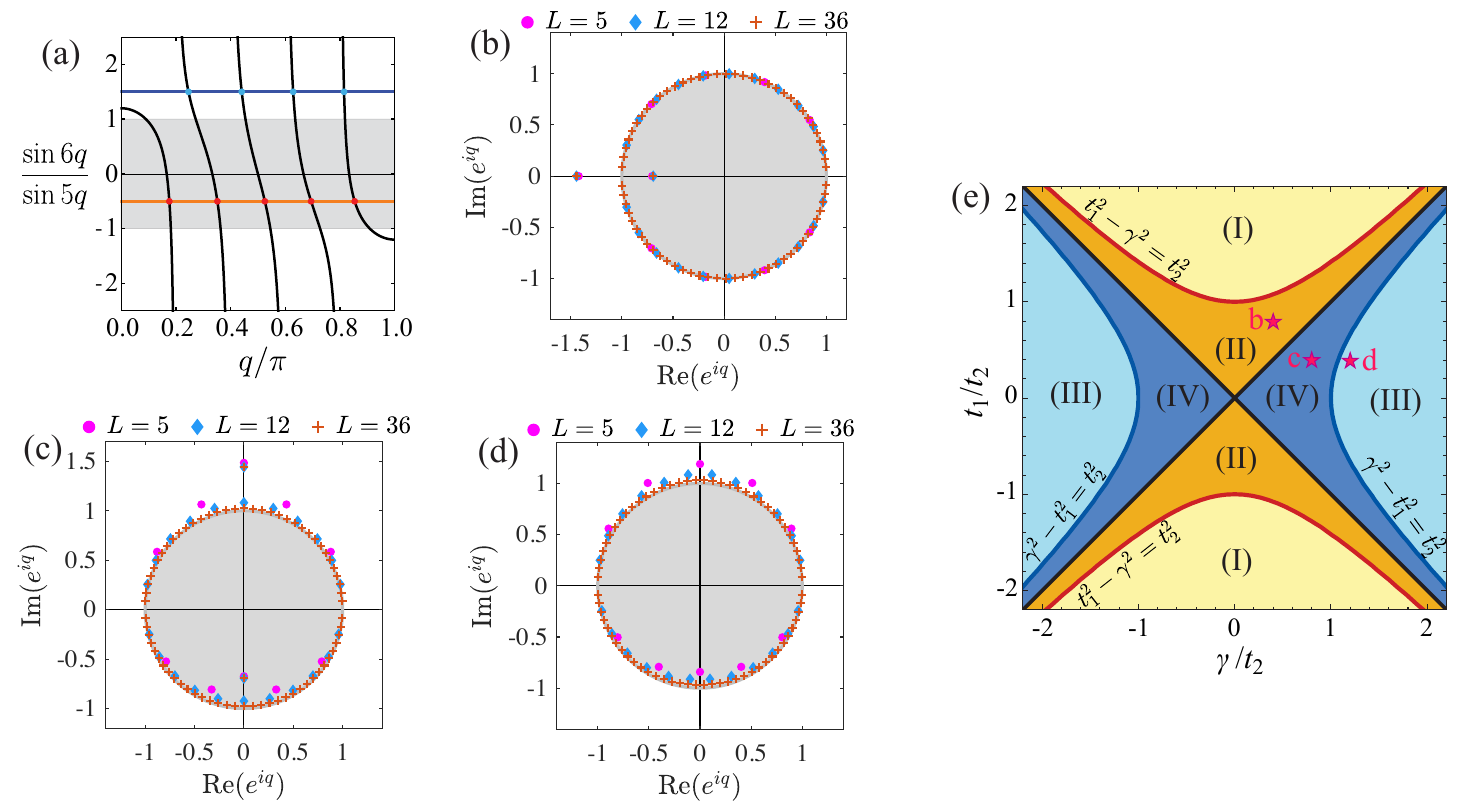}
\caption{Solutions of the generalized wave number $q$ and phase diagram for the nonreciprocal SSH model with intact unit cells. (a) Plot of $\sin [q(L+1)]/\sin (qL)$ with $L=5$. The blue and orange horizontal lines correspond to $y=-t_2/\sqrt{t_1^2-\gamma^2}$ with $t_1=1$, $\gamma=0.6$, and $t_2=-1.2$ and $0.4$, respectively. The gray area between $|y|<1$ marks the topologically trivial phase. (b)-(d) Plots of $e^{iq}$ in the complex plane for solutions of $q$ for different parameter choices and different system sizes: $L = 5$ (magenta circles), $L=12$ (blue diamonds), and $L = 36$ (orange crosses). The gray shaded area is a unit circle added as a guide to the eye. (e)~Phase diagram of the nonreciprocal SSH model. The four different regions are (I) the psH-symmetry unbroken and topologically trivial phase, (II) the psH-symmetry unbroken and topologically nontrivial phase, (III) the psH-symmetry broken and topologically trivial phase, and (IV) the psH-symmetry broken and topologically nontrivial phase. Parameters for (b)-(d) are marked by red stars.}
\label{fig_GBZ}
\end{figure*}

We study a non-Hermitian SSH model with nonreciprocal intracell hoppings $t_1\pm \gamma$ and reciprocal intercell hopping $t_2$ described by the following Hamiltonian
\begin{align}
H = \sum_{n} & \left[(t_1+\gamma) c_{A,n}^{\dagger} c_{B,n}^{\vphantom{\dagger}} +(t_1-\gamma) c^{\dagger}_{B,n} c_{A,n}^{\vphantom{\dagger}} \right. \nonumber\\
&+ \left. t_2 \left(c^{\dagger}_{A, n+1} c_{B,n}^{\vphantom{\dagger}} + c_{B,n}^{\dagger} c_{A,n+1}^{\vphantom{\dagger}}\right)\right],
\label{eq_SSHmodel}
\end{align}
where $c_{\alpha, n}^{\dagger}$ ($c_{\alpha, n}$) is the creation (annihilation) operator in the $n$th unit cell of sublattice $\alpha \in A, B$, hopping parameters $t_1,t_2,\gamma \in \mathbb{R}$, and we assume $t_2>0$ for simplicity throughout the paper. A schematic depiction of this model in the broken and unbroken-unit-cell case can be found in Fig.~\ref{fig_pbc_obc}(a) and (b), respectively. This model preserves chiral symmetry
\begin{align}
C H C^{\dagger} = -H,\quad [C]_{(\alpha\beta|ij)} = \delta_{ij}[\sigma_z]_{\alpha\beta},
\label{eq_chiral_sym}
\end{align}
as well as a pseudo-Hermitian (psH) symmetry
\begin{align}
\eta H\eta^{-1} = H^{\dagger},\quad [\eta]_{(\alpha\beta|ij)} = \delta_{ij}[\sigma_x]_{\alpha\beta}.
\end{align}
The first symmetry dictates that the eigenenergies $E$ come in pairs of opposite sign, i.e., $\{E\} = \{-E\}$, whereas the second symmetry says the eigenenergies come in complex conjugate pairs, i.e., $\{E\} = \{E^*\}$. The Bloch Hamiltonian reads $H(k) = d_x(k) \sigma_x + d_y(k)\sigma_y$, where $d_x(k)=t_1+t_2 \cos k$, $d_y(k)=i\gamma + t_2\sin k$, and $\sigma_{x,y}$ are Pauli matrices. The periodic-boundary-condition~(PBC) spectrum consists of two complex bands
\begin{align} \label{eq:PBC_bands}
E^{\rm PBC}_{\pm}(k) = \pm \sqrt{t_1^2-\gamma^2+t_2^2+2t_2(t_1\cos k+i\gamma \sin k)}.
\end{align}
The disparate spectra under PBCs and OBCs, as shown in Fig.~\ref{fig_pbc_obc}(c-f), indicate the breakdown of the conventional BBC in non-Hermitian systems and point to the NH skin effect~\cite{yao2018edge,yokomizo2019non,kunst2018a}.

For this system, two notions of topology exist. On the one hand, the non-Hermitian point-gap topology is characterized by a quantized spectral winding number about a base point $E_B \in \mathbb{C}$
\begin{align} \label{eq:winding_number}
w\coloneqq \int_{-\pi}^{\pi} \frac{dk}{2\pi i} \partial_k \log\!\;\det [H(k) - E_B],
\end{align}
where a nontrivial winding number $w\neq 0$ signals the piling up of bulk states on the boundaries~\cite{okuma2020topological, zhang2020correspondence}. On the other hand, the band topology is captured by modifying the topological invariant using the generalized Brillouin zone~\cite{yao2018edge,yokomizo2019non,zhang2020correspondence}. For our SSH model this invariant is also a winding number, and its expression is similar to the one in Eq.~\eqref{eq:winding_number} with $E_B = 0$. An alternative approach to quantify the band topology is by using the biorthogonal polarization~\cite{kunst2018a,kunst2019non}, which we also use in this work. Nontrivial values of these invariants point to the appearance of topological boundary states as also exist in Hermitian systems.

In the following we give an overview of the exact solutions of the model in Eq.~\eqref{eq_SSHmodel} with open boundaries in both the broken and unbroken-unit-cell cases~\cite{yao2018edge,kawabata2019prx, kunst2018a, Edvardsson2020Phase}, where we focus solely on the right eigenstates. An analysis of the complete eigensystem of non-Hermitian systems composed of two types of building blocks with OBCs in arbitrary dimensions is given in Ref.~\cite{Yang2025anatomy}.

\subsection{The unbroken-unit-cell case}
\label{sec2a}
We first consider a chain of $L$ intact unit cells as shown in Fig.~\ref{fig_pbc_obc} (a). For an eigenenergy $E\in\mathbb{C}$, the corresponding right eigenstate $|\Psi\rangle = \sum_n(A_n c_{A,n}^{\dagger}+B_n c_{B, n}^{\dagger})|0\rangle$ satisfies the following recurrent relations in the bulk
\begin{align}
t_2 B_{n-1} + (t_1+\gamma) B_n &= E A_n,\nonumber\\
(t_1-\gamma) A_n + t_2 A_{n+1} &= E B_n,\quad n=1,\cdots,L,
\label{eq_SSH_recursion}
\end{align}
and open boundary conditions additionally require
\begin{align}
B_0=A_{L+1}=0. \label{eq_obc}
\end{align}
We take a plane-wave ansatz $A_n \sim A e^{ikn}$, $B_n \sim B e^{ikn} $, where the wave number $k\in \mathbb{C}$ for bulk states.
For $|t_1|\neq|\gamma|$ we can make the substitution~\cite{yao2018edge,kunst2020,okuma2020topological}
\begin{align}
e^{ik} = r e^{iq},\quad r \coloneqq \frac{\sqrt{t_1-\gamma}}{\sqrt{t_1+\gamma}},
\label{eq_similarity_trans}
\end{align}
which is equivalent to a shift of momentum in the complex plane $k=q - i\ln r$, such the bulk equation can be brought to the following symmetrized form
\begin{align}
\begin{pmatrix}
0 & \tilde{t}_1 + t_2 e^{-iq} \\ \tilde{t}_1 + t_2 e^{iq} & 0
\end{pmatrix}
\begin{pmatrix} A_\pm (q) \\ \tilde{B}_\pm (q)
\end{pmatrix} = E_{\pm}^{\rm OBC}(q)
\begin{pmatrix} A_\pm (q) \\ \tilde{B}_\pm (q)
\end{pmatrix}.
\label{eq_SSH_Bloch}
\end{align}
Here $\tilde{t}_1 = \sqrt{t_1^2-\gamma^2}$,  $\tilde{B}_{\pm}=r^{-1}B_{\pm}$, and
\begin{align}
E_{\pm}^{\rm OBC}(q) = \pm \sqrt{t_1^2-\gamma^2+t_2^2 + 2t_2\sqrt{t_1^2-\gamma^2} \cos q}
\label{eq_Eq}
\end{align}
is the OBC band dispersion as a function of $q$. Note that one can obtain this expression for the OBC dispersion by replacing $k = q - i \ln r$ in the PBC dispersion in Eq.~\eqref{eq:PBC_bands}.
For $|\gamma|<|t_1|$, the matrix on the left hand side of Eq.~\eqref{eq_SSH_Bloch} is Hermitian and all eigenenergies are real. This corresponds to the psH-symmetry unbroken phase, where the original NH SSH model in Eq.~\eqref{eq_SSHmodel} can be mapped to a Hermitian one in Eq.~\eqref{eq_SSH_Bloch} with an effective intracell coupling $\tilde{t}_1\in\mathbb{R}$ by a similarity transformation. For $|\gamma|>|t_1|$, however, $\tilde{t}_1$ is purely imaginary and the mapping between the NH SSH model and Hermitian SSH model is broken, also signaled by the fact that $r$ in Eq.~\eqref{eq_similarity_trans} would be imaginary in this case. This corresponds to the psH-symmetry broken phase, where the OBC spectrum is complex. At $|\gamma|=|t_1|$ either $r$ or $r^{-1}$ in Eq.~\eqref{eq_similarity_trans} cannot be defined and the Hamiltonian becomes defective. This is exactly the critical point of the spectral phase transition, and corresponds to the exceptional point of the system under OBCs.

Here we focus on the right eigenstates of the band $E_+^{\rm OBC}(q)$, as those of the other band $E_-^{\rm OBC}(q)$ can be easily obtained by applying the chiral transformation in Eq.~\eqref{eq_chiral_sym}. Noticing the spectral symmetry $E^{\rm OBC}_\pm(q)=E^{\rm OBC}_\pm(-q)$, we write the general solution of a bulk state in the $n$th unit cell as a superposition of $\psi_{n,+}(\pm q) \equiv e^{\pm iqn}(A_+(\pm q), B_+(\pm q))^T$ as
\begin{align}
    \begin{pmatrix} A_{n,+} \\ B_{n,+}
    \end{pmatrix} =\mathcal{N}_q r^n\left[e^{iqn}
    \begin{pmatrix} A_+ (q) \\ B_+ (q)
    \end{pmatrix} + e^{-iqn}
    \begin{pmatrix} A_+ (-q) \\ B_+ (-q)
    \end{pmatrix} \right]
\label{eq_wf}
\end{align}
with $A_{+}(\pm q),B_{+}(\pm q) \in \mathbb{C}$ and $\mathcal{N}_q$ a normalization constant~\cite{kunst2019extended, kunst2020}.  We find $A_+(\pm q)$ and $B_+(\pm q) = r \tilde{B}_+(\pm q)$ from Eq.~\eqref{eq_SSH_Bloch}, such that
\begin{align}
    \frac{B_{+}(\pm q)}{A_{+} (\pm q)} = \frac{\sqrt{ t_1-\gamma}}{\sqrt{t_1+\gamma} }\frac{\sqrt{t_1^2-\gamma^2}+t_2 e^{\pm iq}}{E^{\rm OBC}_+(\pm q)} \equiv C(\pm q).
\label{eq_Cq_def}
\end{align}
The OBCs in Eq.~\eqref{eq_obc} require
\begin{align}
    \begin{pmatrix}
    C(q) & C(-q) \\ e^{iq(L+1)} & e^{-iq(L+1)}
    \end{pmatrix}
    \begin{pmatrix}
    A_{+} (q) \\ A_{+} (-q)
    \end{pmatrix} = 0.
    \label{eq_BCmatrix}
\end{align}
For nontrivial solutions to exist, the determinant of the matrix must vanish, and this leads to the quantization condition of the wave number $q$ as
\begin{align}
    \frac{\sin [q(L+1)]}{\sin (qL)} = - \frac{t_2}{\sqrt{t_1^2 - \gamma^2}}.
\label{eq_q_quantization}
\end{align}

In the psH-symmetry unbroken phase ($|\gamma|<|t_1|$) the right-hand side~(RHS) of the above equation is a real number. As shown in Fig.~\ref{fig_GBZ}(a), there exist $L$ real solutions for $q\in[0,\pi]$ when $|t_2/\tilde{t}_1|<1+L^{-1}$, which in the thermodynamic limit $L\to \infty$ corresponds to the topologically trivial phase $|t_2/\tilde{t}_1|<1$ of the NH SSH model. For $|t_2/\tilde{t}_1|>1+L^{-1}$ there exist only $(L-1)$ real solutions, and the missing solution corresponds to a topological edge mode of complex wave number $q=\pi+iq'$, where $q'\in\mathbb{R}$ is determined by
\begin{align}
\frac{\sinh [q'(L+1)]}{\sinh (q'L)} = \frac{t_2}{\sqrt{t_1^2 - \gamma^2}}.
\label{eq_edgemode_q}
\end{align}
In the thermodynamic limit ($L\to\infty$) one recovers $q' = \ln (t_2/\tilde{t}_1)$ for $q' > 0$, and the energy of the topological edge mode reads
\begin{align}
E_+^\textrm{OBC}(\pi+iq')=\sqrt{\tilde{t}_1^2+t_2^2 - \tilde{t}_1 t_2 (e^{q'}+e^{-q'})} \xrightarrow{L\to \infty} 0 .
\end{align}

In the psH-symmetry broken phase ($|\gamma|>|t_1|$), the RHS of Eq.~(\ref{eq_q_quantization}) is purely imaginary, so that the solutions of $q$ must be found in the complex plane. Again, there exists a topological mode with $q=\pi/2+iq'$, whose energy quickly converges to zero with the increase of system size. Here, $q'\in \mathbb{R}$ is determined by
\begin{align}
\frac{t_2}{\sqrt{\gamma^2-t_1^2}}=
\begin{cases}
\displaystyle -\frac{\sinh[q'(L+1)]}{\cosh(q'L)},& L\in 2\mathbb{Z}^+-1,\\[3mm]
\displaystyle -\frac{\cosh[q'(L+1)]}{\sinh(q'L)},& L\in 2\mathbb{Z}^+,
\end{cases}
\end{align}
such that in the thermodynamic limit we recover $q'=- \ln(t_2/|\tilde{t}_1|)$.

We show numerical solutions of $q$ in the topologically nontrivial phase in the psH-symmetry unbroken and broken case in Figs.~\ref{fig_GBZ}(b) and (c), respectively, and in the topologically trivial phase in the psH-symmetry broken case in Fig.~\ref{fig_GBZ}(d). The phase diagram of the NH SSH model, which comprises four phases depending on whether the psH-symmetry is broken or not, and whether there exist zero-energy topological states or not, is given in Fig.~\ref{fig_GBZ}(e). In the psH-symmetry unbroken phase, the solutions of $q$ for arbitrary system sizes fall exactly on the unit circle $|e^{iq}|=1$, which corresponds to the rescaled generalized Brillouin zone (GBZ)~\cite{yao2018edge}, except for the topological edge modes, cf. Fig.~\ref{fig_GBZ}(b). In the psH-symmetry broken phase, however, the solutions of $q$ in finite-size systems deviate substantially from the GBZ, cf. Figs.~\ref{fig_GBZ}(c) and (d), respectively. The deviation becomes less pronounced as $L$ increases, and will converge to the GBZ when $L\to \infty$. Such system-size dependent spectra in OBC systems have been reported previously~\cite{Gohsrich2024excep}, while the finite-size effect of the GBZ is a unique feature of non-Hermitian systems with complex spectrum and has only been reported recently~\cite{Felski2026parity}.

\subsection{The broken-unit-cell case}
\label{sec2b}
Here we consider a chain with a broken unit cell on the right end as shown in Fig.~\ref{fig_pbc_obc}(b). As such, there are $L+1$ sites on sublattice $A$ and $L$ sites on sublattice $B$, giving $2L+1$ eigenstates in total. The boundary conditions now change to
\begin{align}
B_0 = B_{L+1} = 0,
\end{align}
and consequently Eq.~\eqref{eq_BCmatrix} changes to
\begin{align}
\begin{pmatrix}
1 & 1 \\ e^{iq(L+1)} & e^{-iq(L+1)}
\end{pmatrix}
\begin{pmatrix}
B_{+} (q) \\ B_{+}(-q)
\end{pmatrix} = 0.
\end{align}
For nontrivial solutions to exist the determinant of the matrix on the left-hand side~(LHS) must vanish, which gives the quantization condition of $q$ as
\begin{align}
\sin [q(L+1)] = 0.
\label{eq_q_quantization2}
\end{align}
The generalized wave numbers in the broken-unit-cell case follow a simple expression $q = n \pi/(L+1)$, $n=1, 2,\ldots, L$, which are always real numbers even in the psH-symmetry broken phase. This solution was already previously derived in Ref.~\cite{kunst2020}, and is shown here for the sake of completion.

In addition, there always exists an exact zero-energy mode with vanishing wave function weights on sublattice $B$ due to destructive interference \cite{Kunst2017,Kunst2018,Kunst2019,kunst2018a,kunst2019a} with the right eigenstate 
\begin{equation}
|\Psi_R\rangle = \mathcal{N}_R \sum_{n=1}^L r_R ^n c_{A,n}^{\dagger} |0\rangle, \quad r_R = -\left(\frac{t_1-\gamma}{t_2}\right),\label{eq_exact_zero_mode}
\end{equation}
where $\mathcal{N}_R$ is the normalization factor. We note that in the thermodynamic limit, the topological edge modes in the unbroken-unit-cell case discussed previously in Sect.~\ref{sec2a} can also be written in this form with a partner state localized on the $B$ sublattices on the other end of the chain. Away from the thermodynamic limit this is an approximately good solution as long as the state is well localized.

\subsection{The skin modes and zero modes}
\label{sec2c}

The NH SSH model showcases the NH skin effect, where ``bulk" states are exponentially localized at a certain end of the system under OBCs. In this paper we refer to the states, whose energy are in the OBC spectral continuum, as skin modes, and the zero-energy topological states protected by chiral symmetry as topological modes. We summarize the exact solutions of the right eigenstates derived previously in both the unbroken- and broken-unit cell-cases. From Eqs.~\eqref{eq_Cq_def} and the boundary condition on the left end $B_0=0$ we get
\begin{align}
\frac{B_+(\pm q)}{A_+(\pm q)} = C(\pm q),\qquad \frac{A_+(-q)}{A_+(q)}=-\frac{C(q)}{C(-q)}.
\end{align}
Substituting the coefficients back into Eq.~\eqref{eq_wf}, we obtain the right eigenstates $\Psi_{n,\alpha}(q) \equiv (A_{n,\alpha}(q), B_{n,\alpha}(q))^T$ as
\begin{align}
\Psi_{\alpha,q}(n)=\mathcal{N}_qr^{n}\left[e^{iqn}\psi_{\alpha,q} -e^{-iqn} \psi_{\alpha,-q}\right],
\label{eq_wf_total}
\end{align}
where
\begin{align}
\psi_{n,\alpha}(q) =
\begin{pmatrix}
t_1 + \gamma + r^{-1}t_2e^{-iq} \\ E_{\alpha}^{\rm OBC}(q)
\end{pmatrix},
\label{eq_wf_component}
\end{align}
$\alpha=\pm$ is a label for the two bands as defined in Eq.~(\ref{eq_Eq}), and $\mathcal{N}_q$ is a normalization factor. For a finite-size chain, the generalized momentum $q$ must satisfy Eq.~\eqref{eq_q_quantization} in the unbroken-unit-cell case and Eq.~\eqref{eq_q_quantization2} in the broken-unit-cell case, and there always exists an exact zero mode given by Eq.~\eqref{eq_exact_zero_mode} in the latter case.

The bulk solutions of $q$ are guaranteed to be real for the skin modes in (i) the psH-symmetry unbroken phase in the unbroken-unit-cell case and (ii) all phases in the broken-unit-cell case. This means for (i) and (ii) all the exponentially localized skin modes ($|\Psi(n)|\sim e^{n/\xi_{\rm bulk}}$) share the same localization length $\xi_{\rm bulk}=1/\ln |r|$. When $|r|<1$ ($|r|>1$), the skin modes pile up on the left (right) end of the chain. In (iii) the psH-symmetry broken phase in the unbroken-unit-cell case, however, $q$ remains complex for finite $L$. The imaginary part of $q$ adds additional modulation to the exponential envelope of the wave function amplitudes. However, we emphasize that this effect is hardly observable as $\operatorname{Im}q$ decreases with the increase of system size, and is already considerably small for $L\simeq10$.

\begin{figure}
\centering
\includegraphics[width=\linewidth]{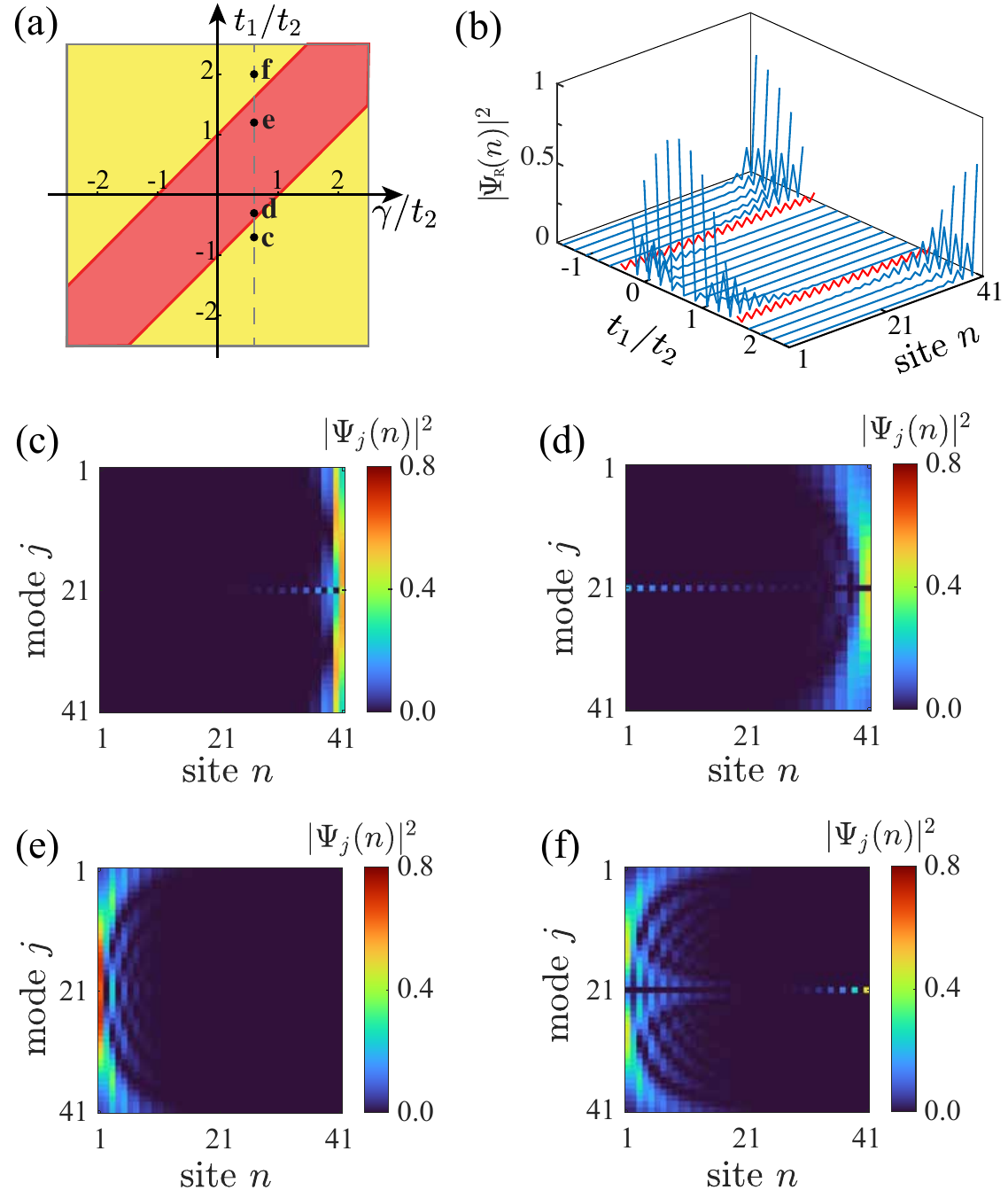}
\caption{Right eigenstates profile of the nonreciprocal SSH model in the broken-unit-cell case. The exact zero mode described by Eq.~(\ref{eq_exact_zero_mode}) is localized on the left (right) end when $|(t_1-\gamma)/t_2|<1$ ($>1$), which corresponds to the red (yellow) region in (a). Its wave function amplitudes as a function of $t_1/t_2$ with $\gamma/t_2=0.6$ and $L=20$ are shown in (b). Wave function amplitudes for all right eigenstates with $\gamma/t_2=0.6$ and $t_1/t_2=-0.7,\ -0.3,\ 1.2$ and $2$ are shown in (c)-(f), respectively. The parameters corresponds to the black dots c-f in (a). The index of modes are sorted in ascending order of the real part of energy.}
\label{fig:SSH_wfa}
\end{figure}

The exact zero mode in Eq.~\eqref{eq_exact_zero_mode} can behave very differently from the skin modes in the broken-unit-cell case. While all the skin modes are localized on the left (right) edge when $\gamma t_1>0$ ($<0$), the exact zero mode, according to Eq.~\eqref{eq_exact_zero_mode}, is localized on the left end for $|(t_1-\gamma)/t_2|<1$ and on the right end for $|(t_1-\gamma)/t_2|>1$, as indicated by the red and yellow regions in Fig.~\ref{fig:SSH_wfa}(a), respectively.  At $|(t_1-\gamma)/t_2|=1$, it delocalizes and becomes an extended state outside a localized continuum, enabling applications such as designing the profile of topological modes by non-Hermiticity~\cite{zhu2021delocalization,wang2022morphing}. In Fig.~\ref{fig:SSH_wfa}(b) we show the wave function profile of the exact zero mode at multiple values of $t_1/t_2$ with $\gamma/t_2=0.6$ fixed. Wave functions for the entire spectrum at selected values of $t_1/t_2$ are shown in Fig.~\ref{fig:SSH_wfa}(c-f). In Fig.~\ref{fig:SSH_wfa}(d) and (f) we can clearly observe that the exact zero mode is localized at the opposite edge of the skin modes.

\begin{figure}
    \centering
    \includegraphics[width=\linewidth]{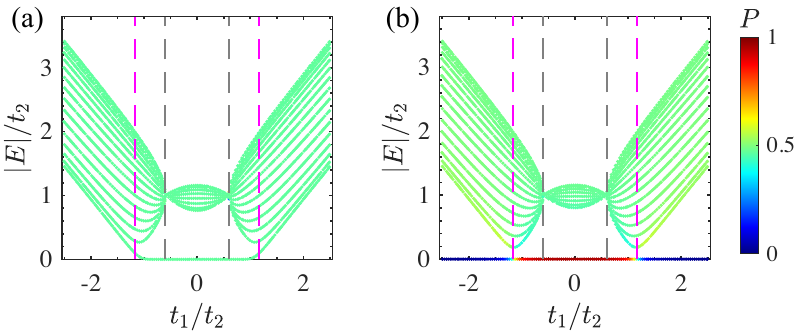}
    \caption{Spectrum of the NH SSH model as a function of $t_1/t_2$ in the (a) unbroken-unit-cell case and (b) broken-unit-cell-case. We fix $t_2=1$ and $\gamma=0.6$. The color of data points show the biorthogonal polarization $P$ as defined in Eq.~\eqref{eq_biortho_polarization} of the corresponding eigenstates. Note that $P$ is calculated numerically for a chain of $16$ unit cells. The gray dashed lines at $t_1=\pm\gamma$ mark the real-to-complex spectral transitions, where psH-symmetry is spontaneously broken and exceptional points with an order scaling with system size appear, and the magenta dashed lines at $t_1=\pm\sqrt{t_2^2+\gamma^2}$ mark the topological phase transitions.}
    \label{fig:biortho_polarization_single_chain}
\end{figure}

Let us make some additional comments. Firstly, we focused only on the right eigenstates so far. To find the left eigenstates, we would have to solve the same equations as before but for $H^\dagger$ instead of $H$. In our specific model, this amounts to changing the sign of $\gamma$. As such, the transpose of left eigenstates $(\langle\Psi_L|)^T = |\Psi_L\rangle$ corresponds to the right eigenstates with $\gamma \rightarrow -\gamma$.

Secondly, the localization of the skin modes and zero modes is solely studied from the perspective of the right eigenstates in this paper. In fact, the skin modes show signatures from ordinary bulk modes if instead their biorthogonal localization, i.e., the projection of the biorthogonal inner product on each sublattice site, would be studied~\cite{Gohsrich2025}. In contrast, the topological edge modes in the unbroken-unit-cell case as well as the zero-energy state in the broken-unit-cell case would still be localized to the boundaries in this picture as long as
\begin{equation}
    |\tilde{t}_1/t_2| =\begin{cases}
        \neq 1, \quad \textrm{broken unit cell}, \\
        < 1, \quad \textrm{unbroken unit cell},
    \end{cases} \label{eq:top_phase_ssh}
\end{equation}
where these expressions can be derived using either the non-Bloch picture~\cite{yao2018edge} or the biorthogonal polarization~\cite{kunst2018a}. This polarization is defined as
\begin{align}
    P\equiv 1-\lim_{L\to\infty}\frac{\langle\psi_L|\sum_{n=1}^{L} n\Pi_n|\psi_R\rangle}{L},
    \label{eq_biortho_polarization}
\end{align}
where $|\psi_R\rangle$ and $|\psi_L\rangle$ are the right and left eigenstate of interest, and $\Pi_n = c^{\dagger}_{A,n} \ket{0} \bra{0}c_{A,n}^{\vphantom{\dagger}} + c^{\dagger}_{B,n} \ket{0} \bra{0}c_{B,n}^{\vphantom{\dagger}}$ is the projector onto each sublattice site in unit cell $n$. When computing this quantity for the zero-energy solution $|\Psi_R\rangle$ in Eq.~\eqref{eq_exact_zero_mode}, where the corresponding left eigenstate $|\Psi_L\rangle$ is given by taking $\gamma \rightarrow - \gamma$ in Eq.~\eqref{eq_exact_zero_mode}, and the normalization of both eigenstates satisfies the biorthogonal condition $\mathcal{N}_L^* \mathcal{N}_R = (r_L r_R)^{-1} (r_L r_R-1)/[(r_L r_R)^{L} - 1]$ with $\mathcal{N}_L$ the normalization coefficient of $|\Psi_L\rangle$, one finds that it takes the quantized values $1$ and $0$, when the state is localized to the right ($|\tilde{t}_1/t_2|<1$) and left ($|\tilde{t}_1/t_2|>1$) end, respectively~\cite{kunst2018a}. Given the considerations mentioned below Eq.~\eqref{eq_exact_zero_mode}, this means that in the unbroken-unit-cell case, $P = 1$ $(0)$ marks the presence (absence) of topological edge modes~\cite{kunst2018a}. 

We show the spectrum of the NH SSH model in the unbroken- and broken-unit-cell case in Figs.~\ref{fig:biortho_polarization_single_chain}(a) and (b), respectively, and color the bands with the value of the associated biorthogonal polarization. Note that while $P$ is defined in the limit of $L\to\infinity$, our results are obtained from numerical calculations of a chain with $16$ full unit cells.
In the broken-unit-cell case, cf. Fig.~\ref{fig:biortho_polarization_single_chain}(b), we see that the values of $t_1/t_2$ for which $P$ of the exact zero mode changes from $0$ to $1$ coincide with the topological phase transitions indicated by the magenta dashed line. The remaining bands correspond to skin modes, and their biorthogonal polarization is always $P=1/2$ regardless of system parameters.
In the unbroken-unit-cell case, as shown in Fig.~\ref{fig:biortho_polarization_single_chain}(a), $P\to 1/2$ holds true for all states, including the topological edge modes whose biorthogonal projections are localized on both ends. Indeed, in this case superpositions of the zero mode localized to the $A$ sublattices $|\Psi_{R}\rangle$ as in Eq.~\eqref{eq_exact_zero_mode} and its partner solution on the $B$ sublattices $|\Psi_{R,B}\rangle$, i.e., $\sim (|\Psi_{R}\rangle \pm |\Psi_{R,B}\rangle)/\sqrt{2}$, are solutions inside the region $|\tilde{t}_1/t_2|<1$. For these superpositions, $P\to 1/2$ as recovered by the numerics. We still see, however, that the values of $P$ as computed for the zero mode in the broken-unit-cell case accurately capture the existence of zero-energy topological modes in the unbroken-unit-cell case. We refer the readers to Appendix \ref{appendix} for analytical derivations of $P$ for the skin modes and its power-law convergence towards the value $1/2$. 
In the following, we will mainly focus on the right eigenstates alone, while keeping these observations in mind.

\begin{figure*}
\centering
\includegraphics[width=0.9\linewidth]{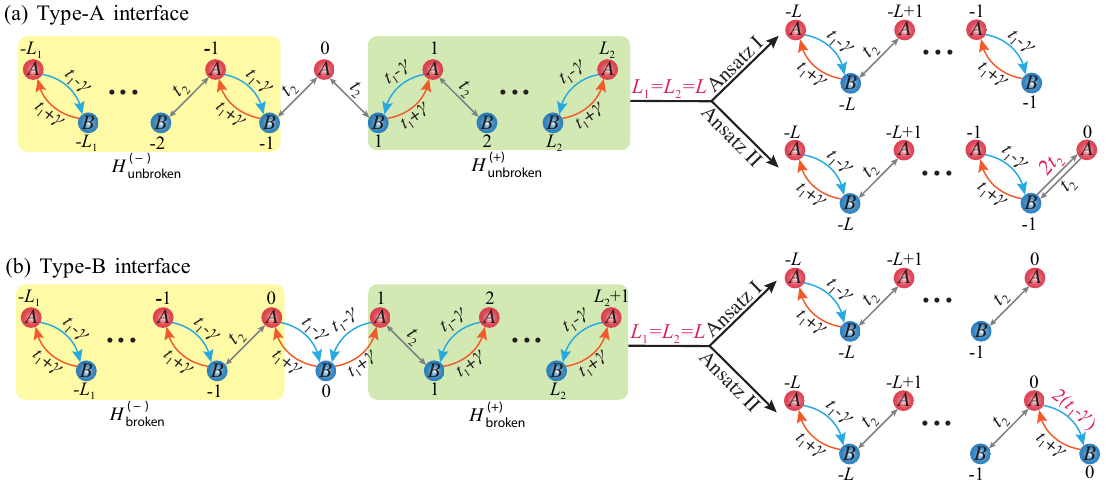}
\caption{The nonreciprocal SSH model with (a) type-A interface and (b) type-B interface in the middle.}
\label{fig_domain_wall}
\end{figure*}

\section{Exact solutions for 1D systems with domain walls}
\label{sec3}

In this section, we present the exact solutions of NH SSH systems with a domain wall in the middle. We consider two types of domain-wall interfaces as shown in Fig.~\ref{fig_domain_wall}: the type-A interface with an $A$ site connected to the left and right blocks by the intercell couplings $t_2$, cf. Fig.~\ref{fig_domain_wall}(a), and the type-B interface with a $B$ site connected to the left and right blocks by the intracell couplings $t_1\pm\gamma$, cf. Fig.~\ref{fig_domain_wall}(b). The left~(-) and right (+) blocks share the same hopping parameters with the direction of intercell and intracell hoppings inverted, i.e., with opposite sign in $\gamma$. This leads to a mirror (inversion) symmetry when the lengths of the two blocks are equal ($L_1=L_2$). In the following we propose two ansatzes of the wave functions for the right eigenstates, and show how each ansatz can be solved exactly using the methods presented in the previous section.

The eigenstates of the system with a type-A interface can be written as
\begin{align}
\Psi^{\text{type-A}} =
\begin{pmatrix}
\Psi^{(-)}_{\rm unbroken} \\ \psi_{A,0} \\ \Psi^{(+)}_{\rm unbroken}
\end{pmatrix},
\end{align}
where $\Psi^{(-)}_{\rm unbroken}=(\psi_{A,-L_1},\psi_{B,-L_1},\cdots,\psi_{A,-1},\psi_{B,-1})^T$ is the wave function of the left block with $L_{1}$ full unit cells, and $\Psi^{(+)}_{\rm unbroken}=(\psi_{B,1},\psi_{A,1},\cdots,\psi_{B,L_2},\psi_{A,L_2})^T$ is the wave function of the right block with $L_{2}$ full unit cells. Similarly, the eigenstates of the system with a type-B interface can be written as
\begin{align}
\Psi^{\text{type-B}} =
\begin{pmatrix}
\Psi^{(-)}_{\rm broken} \\ \psi_{B,0} \\ \Psi^{(+)}_{\rm broken}
\end{pmatrix},
\end{align}
where $\Psi^{(-)}_{\rm broken}=(\psi_{A,-L_1},\psi_{B,-L_1},\cdots,\psi_{B,-1},\psi_{A,0})^T$ is the wave function of the left block with $L_{1}$ full unit cells and a broken unit cell on the right end, and $\Psi^{(+)}_{\rm broken}=(\psi_{A,1},\psi_{B,1},\cdots,\psi_{B,L_2},\psi_{A,L_2+1})^T$ is the wave function of the right block with $L_{2}$ full unit cells and a broken unit cell on the left end. In the following, we always assume $L_1 = L_2 = L$.

\subsection{Type-A interface}
We start by first considering systems with a type-A interface. The Hamiltonian reads
\begin{align}
H^{\text{type-A}} =
\begin{pmatrix}
H_{\rm unbroken}^{(-)} & v & \\
v^{T} & 0 & w^{T} \\ & w& H_{\rm unbroken}^{(+)}
\end{pmatrix},
\end{align}
where $v = (0, \cdots, 0,t_2)^T$ is a $2L$-component vector, $w=\Lambda v$, $H^{(+)}_{\rm unbroken}$ and $H^{(-)}_{\rm unbroken}$ are the Hamiltonians on the left and right of the interface, and $H^{(+)}_{\rm unbroken} = \Lambda H^{(-)}_{\rm unbroken} \Lambda^{\dagger} = (H^{(-)}_{\rm unbroken})^\dagger$. Here
\begin{align}
\Lambda =
\begin{pmatrix}&&&1\\&&1\\&.{}^{\,\displaystyle{}.}{}^{\,\displaystyle{}^{\displaystyle{}.}}\\1
\end{pmatrix}
\label{eq_mirror}
\end{align}
is a unitary transformation ensuring that the spectra of $H^{(+)}_{\rm unbroken}$ and $H^{(-)}_{\rm unbroken}$ are identical.
Due to the mirror (inversion) symmetry about the interface, we can write two ansatzes for the wave functions, where ansatz I reads
\begin{align}
\Psi^{\text{type-A}}_{\rm I} =
\begin{pmatrix}
\Psi^{(-)}_{\rm unbroken, I} \\ 0 \\ -\Psi^{(+)}_{\rm unbroken,I}
\end{pmatrix}, \label{eq:type_a_ansatz_I}
\end{align}
and ansatz II reads
\begin{align}
\Psi^{\text{type-A}}_{\rm II} =
\begin{pmatrix}
\Psi^{(-)}_{\rm unbroken, II} \\ \psi_{\mathrm{II},A,0}\neq 0 \\ \Psi^{(+)}_{\rm unbroken,II}
\end{pmatrix}.
\end{align}
For both ansatzes the wave functions on the left and right blocks are related by the transformation 
\begin{equation}
\Psi^{(+)}_{\rm unbroken, I/II} = \Lambda \Psi^{(-)}_{\rm unbroken, I/II} \,. \label{eq:transformation_left_right_block_interface_A}
\end{equation}

For ansatz I, the left and right blocks are essentially decoupled due to the vanishing amplitude $\psi_{A,0}$. Therefore, $\Psi^{(\mp)}_{\rm unbroken,I}$ should be eigenstates of $H^{(\mp)}_{\rm unbroken}$, and the exact solutions are given in Sec.~\ref{sec2a}. We note that the relative minus sign between $\Psi^{(\mp)}_{\rm unbroken,I}$ in Eq.~\eqref{eq:type_a_ansatz_I} is necessary to ensure $\psi_{\mathrm{II},A,0} = 0$. We further comment that ansatz I gives rise to $2L$ eigenstates and eigenenergies that fully replicate those of the nonreciprocal SSH model with OBCs and unbroken unit cells.

The remaining $2L+1$ solutions have the form of ansatz II. In this case, the non-zero $\psi_{\mathrm{II},A,0}$ introduces effective coupling between the left and right blocks. Due to the mirror (reflection) symmetry, we can still reduce the problem of solving the whole system to solving for a smaller subsystem, which includes either the left or right block plus the interface. Here we focus on the left block and use the same plane-wave ansatz as in Eq.~\eqref{eq_wf}. The open boundary condition on the left end reads
\begin{equation}\psi_{\text{II},B,-(L+1)}=0, \label{eq:type_A_sol_II_boundary_condition_end}
\end{equation}
while the boundary condition at the interface needs to be modified to
\begin{align}
E^{\rm OBC} \psi_{\text{II},A,0} &= t_2 \psi_{\text{II},B,-1} + t_2 \psi_{\text{II},B,+1} \nonumber \\
&= 2t_2 \psi_{\text{II},B,-1}, \label{eq:type_A_sol_II_boundary_condition_int}
\end{align}
where we use $\psi_{\text{II},B,-1}=\psi_{\text{II},B,+1}$ due to the relation in Eq.~\eqref{eq:transformation_left_right_block_interface_A}.
This leads to an effective Hamiltonian
\begin{align}
H_{\rm eff} &=
\begin{pmatrix}
0 & t_1+\gamma &&&&& \\ t_1-\gamma & 0 & t_2 &&&& \\ & t_2 & 0 & \ddots &&& \\
&& \ddots & \ddots&&& \\ &&&& 0 & t_1+\gamma& \\
&&&& t_1-\gamma & 0 & t_2 \\
&&&& & 2t_2 & 0
\end{pmatrix},
\end{align}
for the part of the wave function
\begin{align}
\Psi=(\psi_{\text{II},A,-L},\psi_{\text{II},B,-L},\cdots,\psi_{\text{II},B,-1},\psi_{\text{II},A,0})^T
\end{align}
that is decoupled from the right block. Inserting the plane-wave ansatz from Eq.~\eqref{eq_wf} into the boundary conditions \eqref{eq:type_A_sol_II_boundary_condition_end} and \eqref{eq:type_A_sol_II_boundary_condition_int}, we find
\begin{align}
&B_+ (q) e^{-iq(L+1)}+B_+ (-q) e^{iq(L+1)} = 0, \nonumber \\
&2t_2[B_+ (q) e^{-iq}+B_+ (-q) e^{iq}] \nonumber \\
&\quad - rE^{\rm OBC}_+(q)[A_+ (q) + A_+(-q)]=0,
\end{align}
or equivalently
\begin{align}
\begin{pmatrix}
e^{-iq(L+1)} & e^{iq(L+1)} \\ \sqrt{t_1^2 - \gamma^2} -t_2 e^{-iq} & \sqrt{t_1^2 - \gamma^2} -t_2 e^{iq}
\end{pmatrix}
\begin{pmatrix}
B_+ (q) \\ B_+(-q)
\end{pmatrix} = 0,
\end{align}
where we used Eqs.~\eqref{eq_Eq} and \eqref{eq_Cq_def}.
For nontrivial solutions to exist the determinant of the matrix on the LHS must vanish, and this leads to the following quantization condition of the wave number $q$
\begin{align}
\frac{\sin q(L+1)}{\sin qL} = \frac{t_2}{\sqrt{t_1^2 -\gamma^2}},
\label{eq_additional_mode}
\end{align}
which differs from the quantization condition of ansatz I by a minus sign in the RHS, cf. Eq.~\eqref{eq_q_quantization}.

In the psH-symmetry-unbroken topological phase of the nonreciprocal SSH model, both ansatz I and II comprise of $2L-2$ skin modes with $q\in\mathbb{R}$ and two edge modes with $q\in\mathbb{C}$. However,
instead of giving rise to (nearly) zero-energy states with a wave number $q=\pi+iq'$, where $q'$ is determined by  Eq.~\eqref{eq_edgemode_q}, the edge modes from ansatz II have a complex wave number $q=iq'$ and give additional energy levels outside the spectral continuum
\begin{align}
E_{\pm}^{\rm OBC}(iq')&=\pm \sqrt{\tilde{t}_1^2+t_2^2+\tilde{t}_1t_2(e^{q'}+e^{-q'})} \nonumber \\
&\xrightarrow{L\to \infty} \pm \sqrt{2(\tilde{t}_1^2+t_2^2)}.
\end{align}
The results also hold in the psH-symmetry-broken topological phase, where the wave number of the additional states changes to $q = -\pi/2 + iq'$.

Meanwhile, we notice that Eq.~\eqref{eq_additional_mode} gives $2L$ solutions while $\operatorname{dim}(H_{\rm eff})=2L+1$. The missing solution is an exact zero mode with vanishing wave function amplitude on all $B$ sites
\begin{align} \label{eq:int_type_A_zero_mode_R}
|\psi_{\text{II},R}\rangle = \mathcal{N}_R\sum_{n=-L}^{L} \mathrm{sgn}(n+0^+)r_R^{|n|} c^{\dagger}_{A,n}|0\rangle,
\end{align}
with $r_R = -(t_1 - \gamma)/t_2$, and is localized at the interface when $|r_R|>1$ and on both ends when $|r_R|<1$. The corresponding left eigenstate reads
\begin{align} \label{eq:int_type_A_zero_mode_L}
    |\psi_{\text{II},L}\rangle = \mathcal{N}_L\sum_{n=-L}^{L} \mathrm{sgn}(n+0^+)r_L^{|n|} c^{\dagger}_{A,n}|0\rangle,
\end{align}
with $r_L = -(t_1 + \gamma)/t_2$.
Moving to the biorthogonal picture, we obtain the updated expressions for the biorthogonal localization of the zero mode in Eq.~\eqref{eq:top_phase_ssh} as the following
\begin{equation}
    |\tilde{t}_1/t_2| =\begin{cases}
        > 1, \quad \textrm{localization at domain wall}, \\
        < 1, \quad \textrm{localization on ends}.
    \end{cases}
\end{equation}
This can be easily derived by extending the biorthogonal polarization defined in Eq.~\eqref{eq_biortho_polarization} to the following expression for domain walls
\begin{equation}
    P = 1 - \lim_{L \rightarrow \infty}  \bigg\langle \psi_{L} \bigg| \frac{\sum_{n = -L}^L |n| \Pi_n}{L} \bigg|\psi_{R}\bigg\rangle.
    \label{eq_biotho_polarization_interface}
\end{equation}
Plugging in the solutions for the zero modes in Eqs.~\eqref{eq:int_type_A_zero_mode_R} and \eqref{eq:int_type_A_zero_mode_L}, and requiring $\left<\psi_{\text{II, L}}|\psi_{\text{II},R}\right> = 1$ so that we find $\mathcal{N}_L^*\mathcal{N}_R = (r_L r_R-1)/[2 (r_Lr_R)^{L+1}-r_Lr_R-1]$, we obtain $P = 1$ when $|r_L r_R| < 1$, and $P = 0$ when $|r_L r_R| > 1$. Again $P$ takes a quantized jump at $|r_L r_R|=1$, i.e., $t_1^2-\gamma^2\pm t_2^2 = 0$, signaling a gap closing in the spectrum of the model and a topological phase transition of the underlying NH SSH model. A surprising feature of the type-A domain-wall system is that, while the additional modes emerging in the topologically nontrivial phase are not protected by chiral symmetry, its biorthogonal polarization nevertheless takes a quantized value $1$ as opposed to the skin modes that always shows $P=1/2$. This suggests that the additional modes encodes topological features of the underlying NH systems and its biorthogonal polarization can also serve as a good indicator of topological phase transitions in the NH domain-wall SSH systems.

\subsection{Type-B interface}
The Hamiltonian of the type-B interface system reads
\begin{align}
H^{\text{type-B}} =
\begin{pmatrix}
  H_{\rm broken}^{(-)} & v_+ & \\
  v_{-}^{T} & 0 & w_{-}^{T} \\ & w_+& H_{\rm broken}^{(+)}
\end{pmatrix},
\end{align}
where $v_{\pm}=(0,\cdots,0,t_1\pm\gamma)^T$ are $(2L+1)$-component vectors, $w_{\pm} = \Lambda v_{\pm}$, and $H^{(+)}_{\rm broken} = \Lambda H^{(-)}_{\rm broken} \Lambda^{\dagger}$. The unitary matrix $\Lambda$ is defined in Eq.~\eqref{eq_mirror}. Again we can write down two ansatzes for the wave functions, where ansatz I reads
\begin{align}
\Psi^{\text{type-B}}_{\rm I} =
\begin{pmatrix}
  \Psi^{(-)}_{\rm broken, I} \\ 0 \\ -\Psi^{(+)}_{\rm broken,I}
\end{pmatrix},
\end{align}
and ansatz II reads
\begin{align}
\Psi^{\text{type-B}}_{\rm II} =
\begin{pmatrix}
  \Psi^{(-)}_{\rm broken, II} \\ \psi_{\text{II},B,0}\neq 0 \\ \Psi^{(+)}_{\rm broken,II}
\end{pmatrix}.
\end{align}
The left and right blocks are related by $\Psi^{(+)}_{\rm broken, I/II} = \Lambda \Psi^{(-)}_{\rm broken, I/II}$.

For ansatz I, the left and right blocks are essentially decoupled due to the vanishing $\psi_{B,0}$. Therefore, $\Psi^{(\mp)}_{\rm broken,I}$ should be eigenstates of the block Hamiltonian $H^{(\mp)}_{\rm broken}$, and the exact solutions are given in Sec.~\ref{sec2b}. The reason for the relative minus sign between $\Psi^{(\mp)}_{\rm broken,I}$ is the same as before. Ansatz I gives rise to $2L+1$ eigenstates and eigenenergies that fully replicate those of the NH SSH model with OBCs and broken unit cells. For the exact zero mode arising from Ansatz I, we can again introduce the biorthogonal polarization similar to Eq.~\eqref{eq_biotho_polarization_interface} and its quantized jump from $0$ to $1$ reflects the topological transition.

For ansatz II, we focus on the left block and the interface as in the previous subsection. The open boundary condition on the left end
\begin{equation}
\psi_{\text{II},B,-(L+1)}=0 \label{eq:type_B_sol_II_boundary_condition_end}
\end{equation}
remains intact, whereas the boundary condition at the interface changes to
\begin{align}
E^{\rm OBC} \psi_{\text{II},B,0} &= (t_1-\gamma) \psi_{\text{II},A,0} + (t_1-\gamma) \psi_{\text{II},A,1} \nonumber \\
&= 2(t_1-\gamma) \psi_{\text{II},A,0}, \label{eq:type_B_sol_II_boundary_condition_int}
\end{align}
where we use $\psi_{\text{II},A,0} = \psi_{\text{II},A,1}$ due to $\Psi^{(+)}_{\rm broken, II} = \Lambda \Psi^{(-)}_{\rm broken, II}$.
This leads to an effective Hamiltonian
\begin{align}
H_{\rm eff} &=
\begin{pmatrix}
  0 & t_1+\gamma &&&&& \\ t_1-\gamma & 0 & t_2 &&&& \\ & t_2 & 0 & \ddots &&& \\
  && \ddots & \ddots&&& \\ &&&& 0 & t_2& \\
  &&&& t_2 & 0 & t_1+\gamma \\
  &&&& & 2(t_1-\gamma) & 0
\end{pmatrix},
\end{align}
for the part of the wave living on the left block and at the interface, i.e.,
\begin{align}
\Psi=(\psi_{\text{II},A,-L},\psi_{\text{II},B,-L},\cdots,\psi_{\text{II},A,0},\psi_{\text{II},B,0})^T,
\end{align}
which is decoupled from the right block. Inserting the plane-wave function ansatz Eq.~\eqref{eq_wf} into the boundary conditions in Eqs.~\eqref{eq:type_B_sol_II_boundary_condition_end} and \eqref{eq:type_B_sol_II_boundary_condition_int} we find
\begin{align}
&B_+ (q) e^{-iq(L+1)}+B_+(-q) e^{iq(L+1)} = 0, \nonumber \\
&2(t_1-\gamma)[A_+ (q) +A_+ (-q)] \nonumber \\
&\quad - E^{\rm OBC}_+(q)[B_+ (q) + B_+(-q)]=0,
\end{align}
or equivalently
\begin{align}
\begin{pmatrix}
  e^{-iq(L+1)} & e^{iq(L+1)} \\ \alpha-\beta(e^{iq}-e^{-iq}) & \alpha+\beta(e^{iq}-e^{-iq})
\end{pmatrix}
\begin{pmatrix}
  B_+ (q) \\ B_+ (-q)
\end{pmatrix} = 0,
\end{align}
where we use Eq.~\eqref{eq_Cq_def}, and define $\alpha \equiv t_1^2 -\gamma^2 - t_2^2$ and $\beta \equiv t_2\sqrt{t_1^2-\gamma^2}$.
For nontrivial solutions to exist the determinant of the matrix on the LHS must vanish, and this leads to the quantization condition of the wave number $q$ according to
\begin{align}
\cot[q(L+1)]\sin q = \frac{t_1^2-\gamma^2-t_2^2}{2t_2\sqrt{t_1^2-\gamma^2}}.
\end{align}
One can always find $L$ solutions of $q$ corresponding to the skin modes, which are real in the psH-symmetry-unbroken phase and complex in the psH-symmetry-broken phase. In addition, there always exists a special solution corresponding to either the topological zero mode or finite-energy additional modes outside the spectral continuum of the NH SSH model. To determine the special solution, we consider the thermodynamic limit $L\to \infty$, where the above equation is reduced to
\begin{align}
-i\operatorname{sgn}(\operatorname{Im}q)\sin q = \frac{t_1^2-\gamma^2-t_2^2}{2t_2\sqrt{t_1^2-\gamma^2}}.
\end{align}
The solutions reads
\begin{align}
q =
\begin{cases}
  iq',& \text{PH-unbroken trivial phase},\\
  \pi +iq',& \text{PH-unbroken topological phase},\\
  \pi/2-iq', & \text{PH-broken phases},
\end{cases}
\end{align}
where
\begin{align}
q'=
\begin{cases}
  \displaystyle\sinh^{-1}\left|\frac{t_1^2-\gamma^2-t_2^2}{2t_2\sqrt{t_1^2-\gamma^2}}\right|, & \text{PH-unbroken phase},\\
  \displaystyle\cosh^{-1}\left|\frac{t_1^2-\gamma^2-t_2^2}{2t_2\sqrt{t_1^2-\gamma^2}}\right|, & \text{PH-broken phase}.
\end{cases}
\end{align}
Inserting the solution of $q$ into the dispersion relation under OBCs in Eq.~\eqref{eq_Eq}, we get the energy of the special solution $E^{\rm OBC}(q)=0$ in the topologically nontrivial phases in both the psH-symmetry broken and unbroken regimes, namely $|(t_1^2-\gamma^2)/t_2^2|<1$, and $E^{\rm OBC}(q)=\sqrt{2(t_1^2-\gamma^2 +t_2^2)}$ in the topologically trivial phases in both the psH-symmetry broken and unbroken regimes, namely $|(t_1^2-\gamma^2)/t_2^2|>1$. The latter case corresponds to the additional modes outside the spectral continuum. We notice that the additional modes in type-A and type-B interface systems share the same dispersion, but appear in complementary phase regions. For type-A interfaces, the additional levels always appear in the topological phase of the NH SSH model, together with the topological zero modes; but for type-B interfaces, the additional levels always appear in the trivial phase of the NH SSH model. We note that in certain parameter regimes of the PH-broken trivial phase, the imaginary part of the energies of the additional modes can be larger than the largest imaginary part of the bulk bands, making it the dominant mode in the long-time limit.

We emphasize that the two ansatzes and exact solutions only apply to systems with inversion symmetry, namely $L_1=L_2$. However, the localization properties and spectral properties of all the eigenstates, especially the existence of additional states and their dispersions, remain correct as long as $L_{1,2}$ are both moderately large. In Sec.~\ref{sec4} we will show numerical results for not only 1D but also 2D systems with interfaces of codimensions one and two, further generalizing our conclusions.

\begin{figure}
\centering
\includegraphics[width=\linewidth]{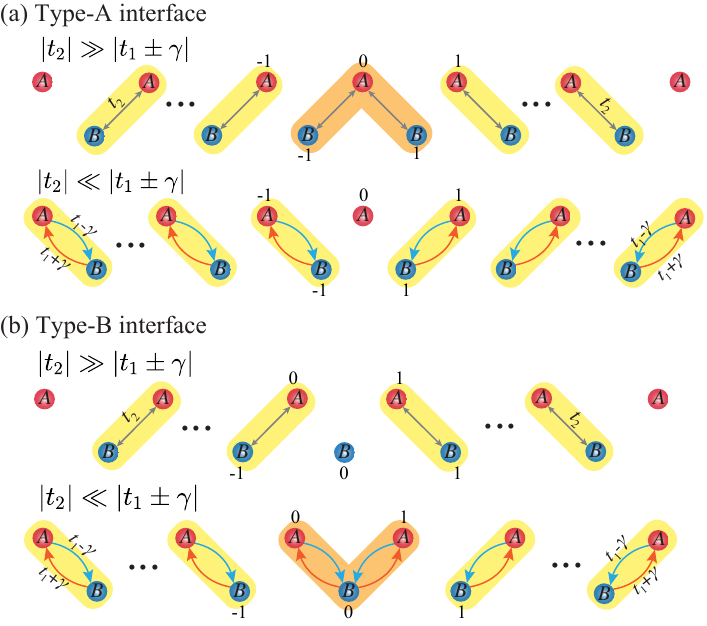}
\caption{Dimerized limits of the NH SSH model with two types of interfaces in (a) and (b). The dimer states are highlight in yellow and the trimer states at the interface are highlighted in orange.}
\label{fig_dimer}
\end{figure}

\subsection{Dimerized limit}

\begin{figure*}[tb]
    \centering
    \includegraphics[width=0.8\linewidth]{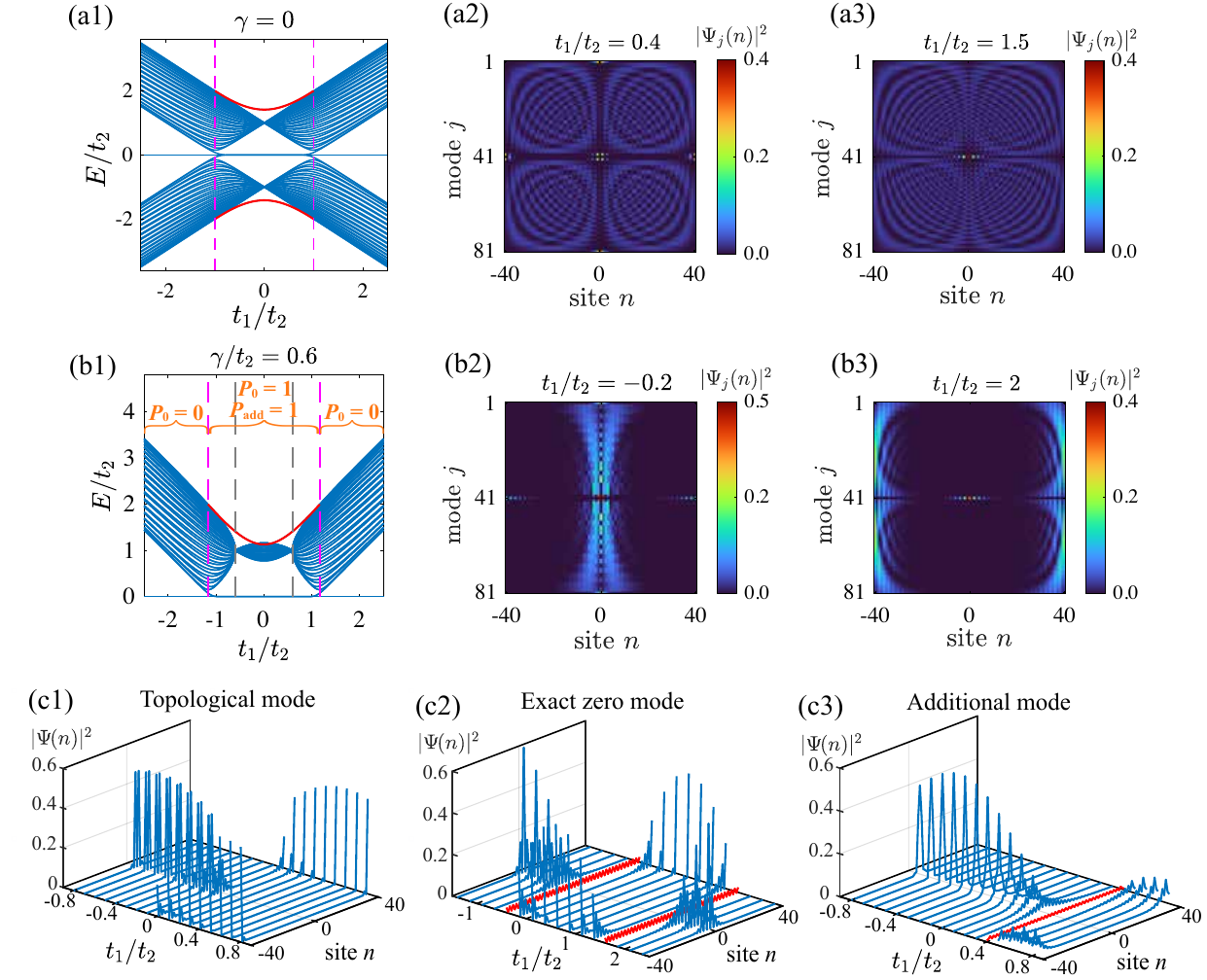}
    \caption{Energy spectrum and wave functions of the type-A interface system with $L_1=L_2=20$. (a1)-(a3) show the real OBC spectrum and representative right-eigenstate profiles in the Hermitian limit $\gamma=0$. (b1)-(b3) show the complex OBC spectrum and representative right-eigenstate profiles for $\gamma/t_2=0.6$. The mode indices are sorted by the real part of energy in ascending order. The magenta dashed lines in (a1) and (b1), and the gray dashed lines in (b1) mark the topological phase transition and psH-symmetry broken transition, respectively. The bands marked in red in (a1) and (b1) are the additional modes outside the continuum bands. In (b1) we additionally show the regions, where the biorthogonal polarization $P_0$ for the exact zero mode and $P_{\rm add}$ for the additional mode take quantized value $0$ or $1$. (c1)-(c3) shows the wave function profiles of the topological mode, the exact zero mode, and the additional mode at various values of $t_1/t_2$, while $\gamma/t_2=0.6$ is fixed. The red curves in (c2) and (c3) highlight that both the zero mode and additional modes become extended outside of the continuum.}
    \label{fig:typeA_1d}
  \end{figure*}

The dimerized limit of the SSH model provides a clear picture for the origin of additional states from interface structures. For a system with a type-A interface, when the intercell hopping is much stronger than intracell hopping ($|t_2|\gg |t_1\pm \gamma|$), the dimers are formed by $A$ and $B$ sites in adjacent unit cells, leaving two single $A$ sites on the left/right ends and a trimer at the interface [see Fig.~\ref{fig_dimer}(a)]. The single sites correspond to zero-energy topological modes. The trimer block is described by the Hamiltonian
\begin{align}
  H_{\rm trimer}^{\text{type-A}} =
  \begin{pmatrix}
    0 & t_2 & 0 \\
    t_2 & 0 & t_2 \\
    0 & t_2 & 0
  \end{pmatrix},
\end{align}
leading to a zero-energy state with nonzero wave function only on the $A$ site formed at the interface, and two states with energies $E=\pm \sqrt{2t_2^2}$ corresponding to the additional states. For finite $t_1\pm \gamma$, the dimer and trimer blocks hybridize and the dispersion of finite energy states is modified. However, the zero modes on the boundaries and the additional modes at the interface remain as long as the system stays in the topological phase. This explains why the additional states always appear together with the topological modes in type-A interface systems.

For a system with type-B interface, the trimer block is formed in the limit of $|t_2| \ll |t_1\pm \gamma|$ instead, where the dimers are formed within each unit cell and zero energy topological modes on the two ends are absent [see Fig.~\ref{fig_dimer}(b)]. The trimer block is described by the Hamiltonian
  \begin{align}
    H_{\rm trimer}^{\rm type-B} =
    \begin{pmatrix}
      0 & t_1+\gamma & 0 \\
      t_1-\gamma & 0 & t_1-\gamma \\
      0 & t_1+\gamma & 0
    \end{pmatrix},
  \end{align}
again leading to a zero energy state and two states of energy $E=\pm \sqrt{2(t_1^2-\gamma^2)}$. In the limit of $|t_2| \gg |t_1\pm \gamma|$ that falls onto the topological phase, the $B$ site at the interface is left unpaired, giving rise to yet another zero mode in addition to the two zero modes that reside on the $A$ sites on the both ends of the chain.

\begin{figure*}[tb]
    \centering
    \includegraphics[width=0.8\linewidth]{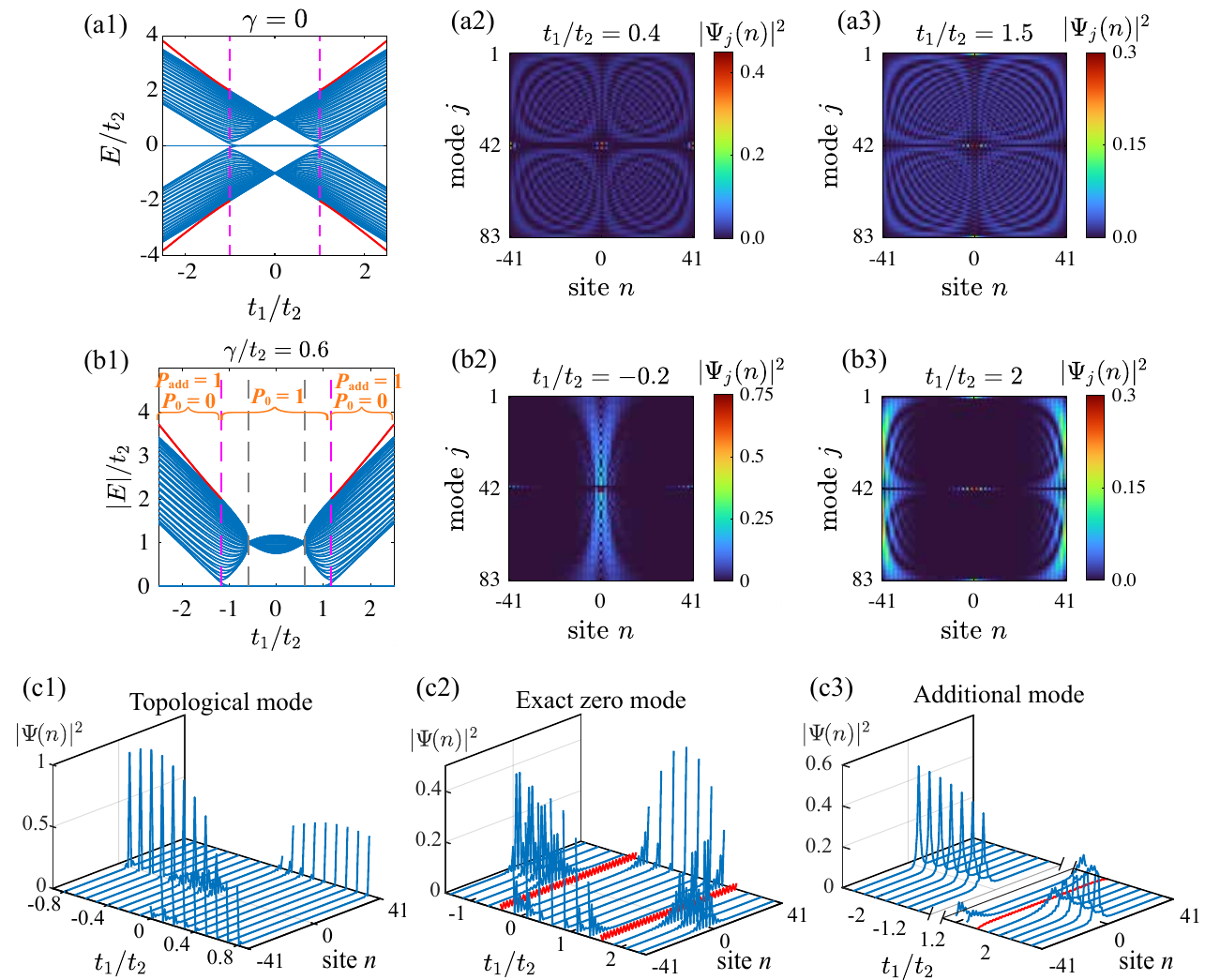}
    \caption{Energy spectrum and wave functions of the type-B interface system with $L_1=L_2=20$. (a1)-(a3) show the real OBC spectrum and representative right-eigenstate profiles in the Hermitian limit $\gamma=0$. (b1)-(b3) show the complex OBC spectrum and representative right-eigenstate profiles for $\gamma/t_2=0.6$. The mode indices are sorted by the real part of energy in ascending order. The dashed magenta lines in (a1) and (b1), and the gray vertical lines in (b1) mark the topological phase transition and psH-symmetry broken transition, respectively. The bands marked in red in (a1) and (b1) are the additional modes outside the continuum bands. In (b1) we additionally show the regions, where the biorthogonal polarization $P_0$ for the exact zero mode and $P_{\rm add}$ for the additional mode take quantized value 0 or 1. (c1)-(c3) shows the wave function profiles of the topological mode, the exact zero mode, and the additional mode at various values of $t_1/t_2$, while $\gamma/t_2=0.6$ is fixed. The red curves in (c2) and (c3) highlight that both the zero mode and additional modes become extended outside of the continuum.}
    \label{fig:typeB_1d}
  \end{figure*}

Finally, we emphasize that the additional modes at finite energy are not protected by chiral symmetry, and are therefore prone to being smeared out by symmetry-preserving hopping disorder. On the other hand, tuning the couplings locally at the domain wall can change the dispersion of the additional modes, while leaving the bulk skin modes largely unaffected. This property endows the domain-wall systems with a greater degree of tunability, and broadens their potential for practical applications.
  
\section{Numerical results}
\label{sec4}


In this section, we present the spectrum and wave functions of the domain-wall systems from exact diagonalization. We focus on the right eigenstates, which directly encode the spatial localization in the presence of an interface. We first show that the numerical results of 1D systems validate the analytical predictions in Sec.~\ref{sec3}. Then we turn to the numerical results of 2D systems and discuss the general conclusions in higher-dimensional setups.

\subsection{One-dimensional systems}
\label{sec4a}
We first benchmark the analytical predictions from Sec.~\ref{sec3} against exact diagonalization, and then use representative right-eigenstate profiles to disentangle the skin, topological, and additional interface modes.

\begin{figure*}
  \centering
  \includegraphics[width=0.7\linewidth]{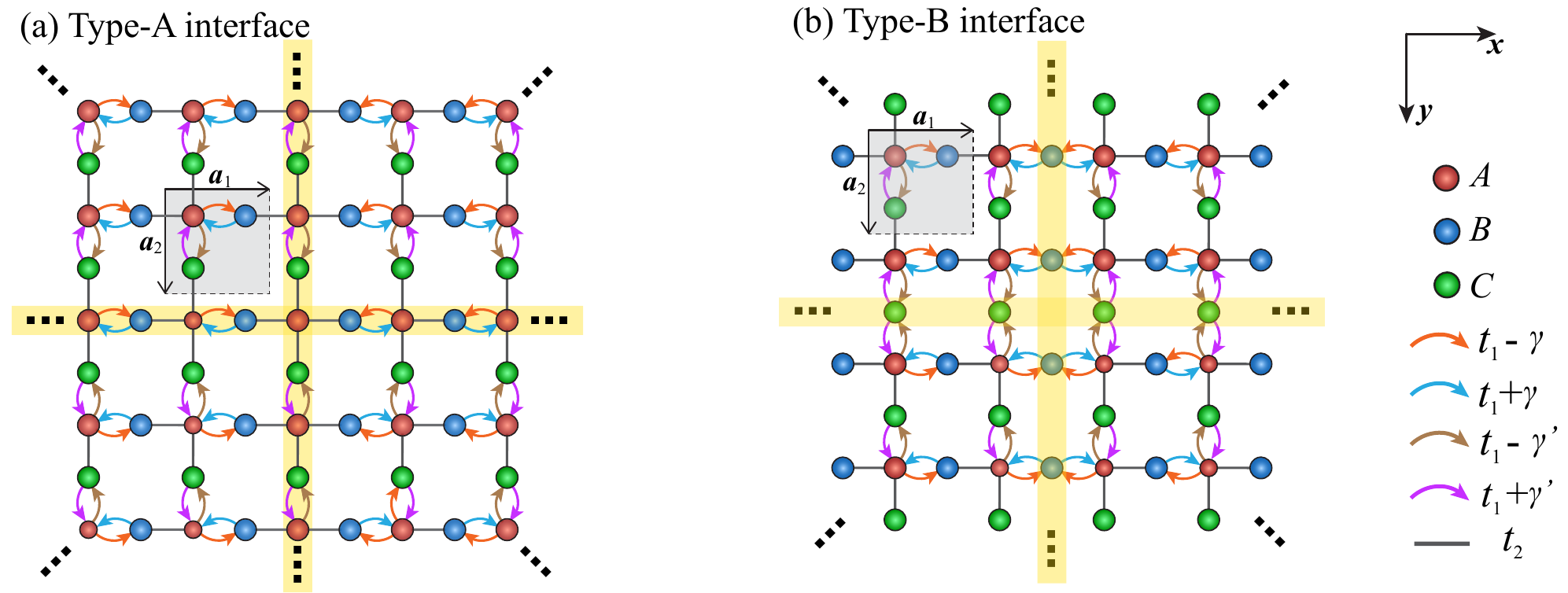}
  \caption{Schematic plot of the two-dimensional Lieb lattice with (a) type-A and (b) type-B interfaces in the middle. $A$, $B$, and $C$ are sites on the three sublattices displayed in red, blue and green, respectively. The codimension-one interfaces are highlighted in the yellow area. The gray rectangle marks the unit cell of the 2D Lieb lattice, with $\boldsymbol{a}_1$ and $\boldsymbol{a}_2$ being the primitive vectors.}
  \label{fig:domain_wall_2D}
\end{figure*}

We start by considering a system with a type-A interface sitting in the middle. In the Hermitian limit $\gamma=0$, as shown in Figs.~\ref{fig:typeA_1d}(a1)--(a3), the spectrum is purely real and exhibits the expected symmetry $E\leftrightarrow -E$. We confirm that the dispersion of the additional modes marked by red lines is consistent with the theoretical prediction $E=\pm\sqrt{2(t_1^2+t_2^2)}$. While the majority of eigenstates are extended over the two segments (bulk states), there always exists a localized eigenstate or eigenstates at any given parameter $t_1/t_2$. The topological phase of the underlying SSH blocks ($|t_1/t_2|<1$) supports five localized states, including one exact zero mode localized on the boundaries, two topological modes with localization centers at both the interface and the boundaries, and two additional modes localized at the interface. In the trivial phase ($|t_1/t_2|>1$) the exact zero mode remains localized but the localization center changes to the interface.


Next we turn on the nonreciprocity, letting $\gamma/t_2=0.6$. Fig.~\ref{fig:typeA_1d}(b1) shows the absolute value of the complex spectrum. 
Again, we observe discrete eigenvalues detached from the bulk continuum (additional modes) in the topological phase, following the dispersion $E=\pm \sqrt{2(t_1^2-\gamma^2+t_2^2)}$ marked by the red line. While $|E|$ of the additional modes may fall inside the bulk continuum for certain parameters, on the complex plane their energies remain well separated. The topological phase transition can be detected by the biorthogonal polarization of the exact zero mode as well as the additional modes. In Fig.~\ref{fig:typeA_1d}(b1) we show the regions, where the biorthogonal polarization $P_0$ for the exact zero mode and $P_{\rm add}$ for the additional mode take quantized value 0 or 1. In the meantime, all the skin modes that lives in the spectral continuum obey $P=1/2$ as discussed in Sec.~\ref{sec2c} and Appendix~\ref{appendix}. Figures.~\ref{fig:typeA_1d}(b2) and (b3) show two cases, where the exact zero mode is localized on different locations as compared to the skin states. Analytical results from previous sections show that the exact zero mode is localized at the interface when $|(t_1-\gamma)/t_2|>1$ and on both ends when $|(t_1-\gamma)/t_2|<1$, while the skin modes are localized at the interface when $\gamma/t_1<0$ and on both ends when $\gamma/t_1>0$.

Finally, we show how the right-eigenstate profiles $|\psi^{\mathrm{R}}_m(j)|^2$ of the topological mode, the exact zero mode, and the additional mode change as $t_1/t_2$ is tuned (with $\gamma/t_2=0.6$ fixed) in Figs.~\ref{fig:typeA_1d}(c1)--(c3), respectively. The topological modes, cf. Figs.~\ref{fig:typeA_1d}(c1), pinned at almost zero energy remain sharply localized throughout the parameter scan inside the topological phase, while the localization center shifts between the interface and the two ends. The additional modes, cf. Figs.~\ref{fig:typeA_1d}(c3), which also exist only in the topological phase, are unique localized states arising from the type-A interface. Noticeably, their energies remain real and isolated from the bulk bands until the onset of the topological transition. Similar to the exact zero mode, cf. Figs.~\ref{fig:typeA_1d}(c2), the additional modes can also become extended at $|(t_1+\gamma)/t_2|=1$ [see the red curves in Fig.~\ref{fig:typeA_1d}(c3)], making it an extended state outside a localized continuum.

Fig.~\ref{fig:typeB_1d} displays the results for a system with a type-B interface in the middle. Again, we consider the Hermitian limit $\gamma=0$, Figs.~\ref{fig:typeB_1d}(a1)-(a3), and a non-Hermitian case $\gamma/t_2=0.6$, Figs.~\ref{fig:typeB_1d}(b1)-(c3). The additional modes now exist in the trivial phase of the underlying SSH blocks, $|\tilde{t}_1/t_2|>1$, and the dispersions $E=\pm\sqrt{2(t_1^2-\gamma^2+t_2^2)}$ remain consistent with the theoretical prediction. The additional modes of type-B interface systems become extended at $|(t_1-\gamma)/t_2|=1$ [see the red curves in Fig.~\ref{fig:typeB_1d}(c3)], again making it an extended state outside a continuum. Moving to the biorthogonal picture, again the topological phase transition can be detected by the biorthogonal polarization of the exact zero mode as well as the additional modes, as shown in Fig.~\ref{fig:typeB_1d}(b1). The difference from the type-A interface is that the additional modes now exist in the topologically trivial phase, and its biorthogonal polarization takes the value $P_{\rm add}=1$ while the biorthogonal polarization for the exact zero mode takes $P_0=1$.

\subsection{Two-dimensional systems}
\label{sec4b}

\begin{figure*}
  \centering
  \includegraphics[width=0.8\linewidth]{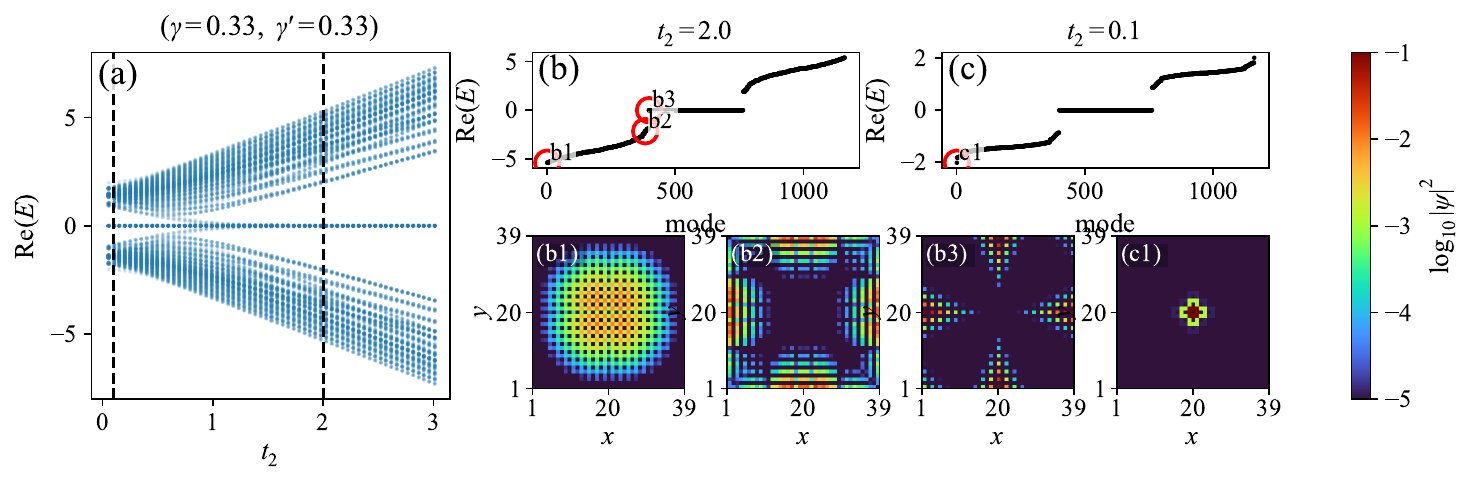}
  \caption{Type-A interface with $(\gamma,\gamma')=(0.33,0.33)$ and $t_1=1$. (a) OBC spectrum as a function of $t_2$; the magenta dashed lines mark the representative cuts at $t_2=0.1$ and $t_2=2$. (b) Spectrum at $t_2=2$ and selected right-eigenstate densities: a bulk skin state funneled to the codimension-two central junction (b1), an edge-localized state (b2), and a corner-localized state (b3). (c)~Spectrum at $t_2=0.1$ featuring a sharply localized additional interface state (c1).}
  \label{fig:2D_center_FULL}
\end{figure*}

\begin{figure*}
  \centering
  \includegraphics[width=0.8\linewidth]{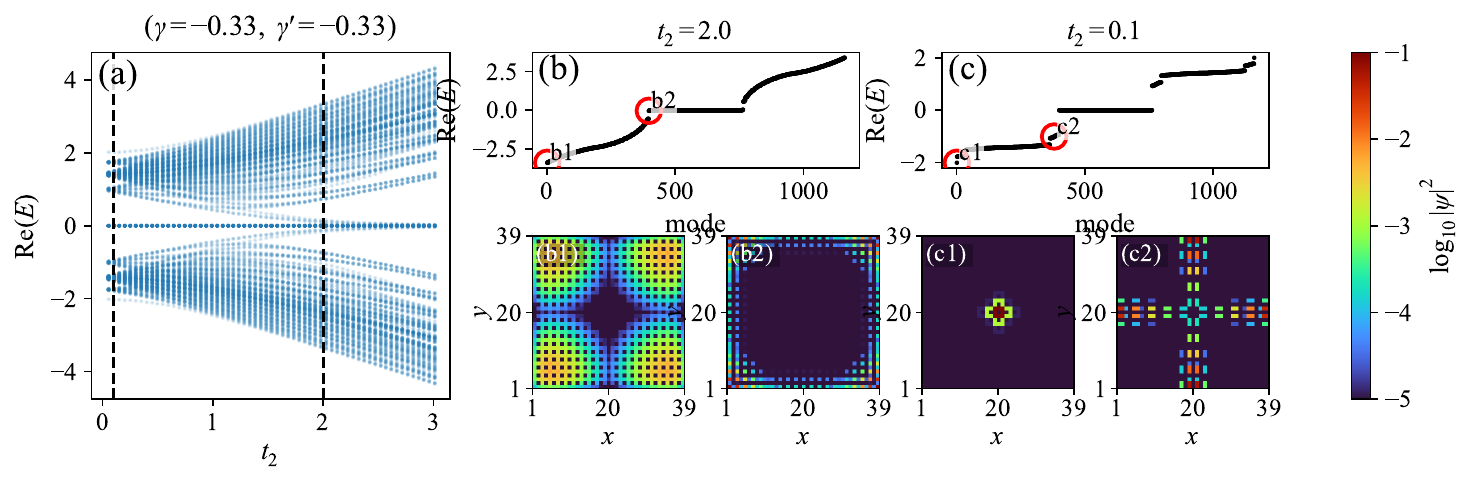}
  \caption{Type-A interface with $(\gamma,\gamma')=(-0.33,-0.33)$ and $t_1=1$. (a) OBC spectrum versus $t_2$, with the representative cuts $t_2=0.1$ and $t_2=2$. (b) Spectrum at $t_2=2$ and representative right-eigenstate densities showing bulk funneling towards the four outer corners (b1) and a corner-localized mode (b2). (c) Spectrum at $t_2=0.1$ together with an additional interface state (c1) and a strongly localized interface mode (c2).}
  \label{fig:2D_corners_FULL}
\end{figure*}

\begin{figure*}
  \centering
  \includegraphics[width=0.8\linewidth]{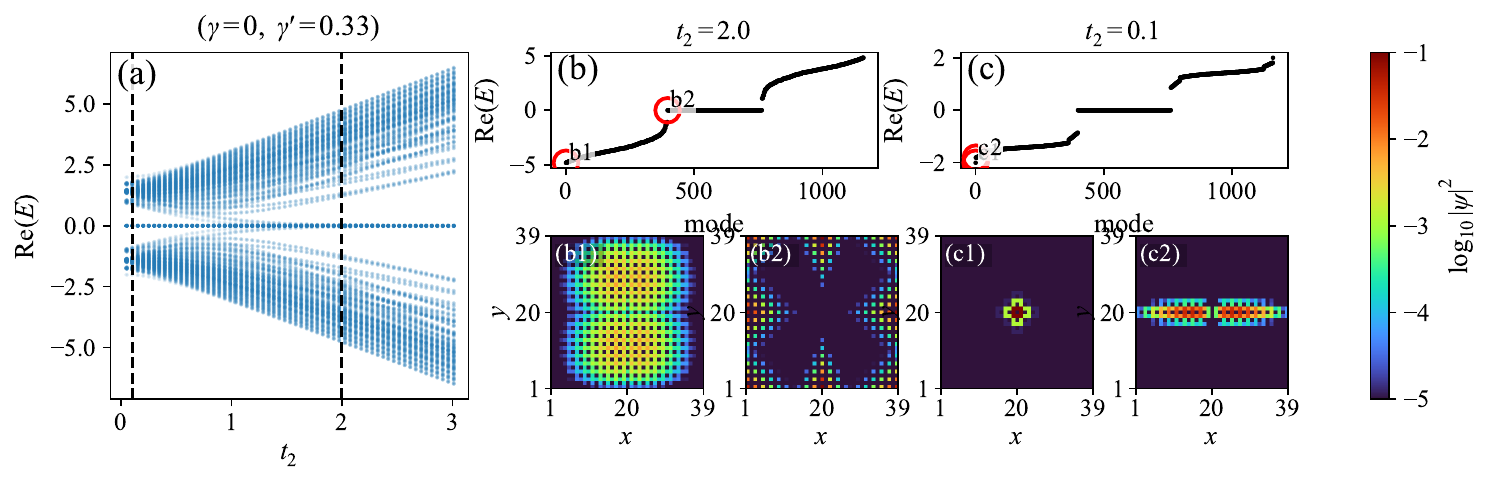}
  \caption{Type-A interface with $(\gamma,\gamma')=(0.00,0.33)$ and $t_1=1$. (a) OBC spectrum versus $t_2$; the magenta dashed lines mark the representative cuts at $t_2=0.1$ and $t_2=2$. (b) Spectrum at $t_2=2$ and representative right-eigenstate densities showing bulk funneling onto the horizontal codimension-one interface (b1) and a corner-localized mode (b2). (c) Spectrum at $t_2=0.1$ together with an additional interface state (c1) and a defect-localized mode on the midline (c2).}
  \label{fig:2D_midline_FULL}
\end{figure*}

\begin{figure*}
  \centering
  \includegraphics[width=0.8\linewidth]{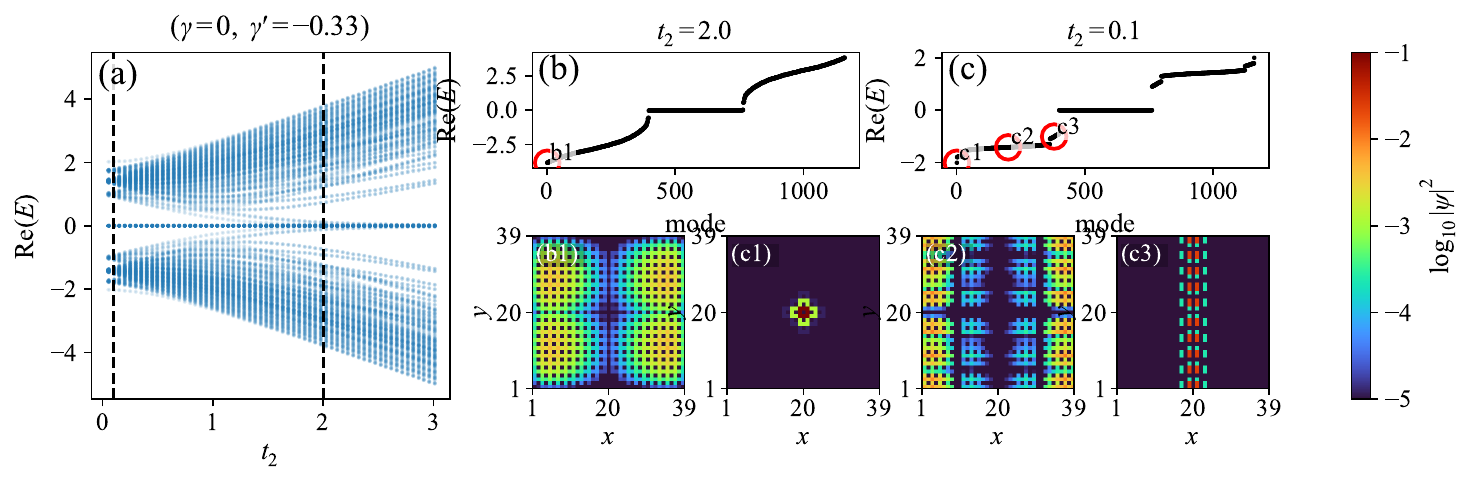}
  \caption{Type-A interface with $(\gamma,\gamma')=(0.00,-0.33)$ and $t_1=1$. (a) OBC spectrum versus $t_2$; the magenta dashed lines again mark the representative cuts at $t_2=0.1$ and $t_2=2$. (b) Spectrum at $t_2=2$ and a representative right-eigenstate density showing bulk funneling towards the two parallel outer edges (b1). (c) Spectrum at $t_2=0.1$ together with an additional interface state (c1), a bulk-funneling state localized at the two outer edges (c2), and a defect-localized state (c3).}
  \label{fig:2D_twoedges_FULL}
\end{figure*}

In this section, we generalize the interface geometry to two dimensions, where the interplay between nonreciprocity and interfaces with codimensions one and two produces distinct funneling regimes that can be diagnosed directly from right-eigenstate density profiles. We extend the nonreciprocal SSH model to a two-dimensional Lieb lattice, cf. Fig.~\ref{fig:domain_wall_2D}. In Figs.~\ref{fig:2D_center_FULL}-\ref{fig:2D_twoedges_FULL}, we fix $t_1=1$ and vary the intercell hopping $t_2$; the representative cuts are taken at $t_2=2$ and $t_2=0.1$. To capture the two independent nonreciprocities used in our interface geometry (Fig.~\ref{fig:domain_wall_2D}), we distinguish the intracell nonreciprocal hoppings on the $A$-$B$ links and the $A$-$C$ links by $\gamma$ and $\gamma'$, respectively. The real-space Hamiltonian reads
\begin{widetext}
  \begin{align}
    H
    =&\sum_{i,j}\Big[(t_1+\gamma)\,c^\dagger_{A,i,j}c_{B,i,j}+(t_1-\gamma)\,c^\dagger_{B,i,j}c_{A,i,j}
    +t_2\big(c^\dagger_{A,i+1,j}c_{B,i,j}+c^\dagger_{B,i,j}c_{A,i+1,j}\big)\Big]\nonumber\\
    &+\sum_{i,j}\Big[(t_1+\gamma')\,c^\dagger_{A,i,j}c_{C,i,j}+(t_1-\gamma')\,c^\dagger_{C,i,j}c_{A,i,j}
    +t_2\big(c^\dagger_{A,i,j+1}c_{C,i,j}+c^\dagger_{C,i,j}c_{A,i,j+1}\big)\Big],
    \label{eq:lieb_realspace}
  \end{align}
\end{widetext}
where $A,B,C$ label the three sublattices, and $B$ ($C$) sits on the $x$- ($y$-) link of each unit cell.
Under PBCs, in the basis $(A,B,C)$ the Bloch Hamiltonian takes the rank-$2$ form
  \begin{equation}
  \begin{aligned}
      &\qquad H(\bm{k})=
    \begin{pmatrix}
      0 & f_x(\bm{k}) & f_y(\bm{k})\\
      g_x(\bm{k}) & 0 & 0\\
      g_y(\bm{k}) & 0 & 0
    \end{pmatrix},
    \\
    &\begin{cases}
      f_x=(t_1+\gamma)+t_2e^{-ik_x},\ \ g_x=(t_1-\gamma)+t_2e^{ik_x},\\[2pt]
      f_y=(t_1+\gamma')+t_2e^{-ik_y},\ g_y=(t_1-\gamma')+t_2e^{ik_y}.
    \end{cases}
  \end{aligned}
    \label{eq:bloch_3band}
  \end{equation}

This immediately yields one flat band at $E_0(\bm{k})=0$, and two dispersive bands
\begin{equation}
\begin{aligned}
&E_\pm^{\rm PBC}(\bm{k})
=\pm\left\{
2(t_1^2+t_2^2)-\gamma^2-(\gamma')^2 \right. \\
&\left. +2t_2[t_1(\cos k_x+\cos k_y)
+i(\gamma\sin k_x+\gamma'\sin k_y)]
\right\}^{1/2}.
\end{aligned}
\label{eq:lieb_pbc_expanded}
\end{equation}
We next turn to open boundaries and the domain-wall configuration in Fig.~\ref{fig:domain_wall_2D}, where the codimension-one interfaces enable the funneling phenomenology shown in Figs.~\ref{fig:2D_center_FULL}-\ref{fig:2D_twoedges_FULL}.

To elucidate the spectral properties and the spatial distribution of eigenstates under open boundary conditions, we must account for the non-Hermitian skin effect. Unlike the extended PBC eigenmodes, whose energies are given by Eq.~\eqref{eq:lieb_pbc_expanded}, the bulk eigenstates in the non-reciprocal system tend to localize exponentially at the boundaries or domain interfaces. In the domain-wall geometry of Fig.~\ref{fig:domain_wall_2D}, the sign pattern of $(\gamma,\gamma')$ fixes the direction of pumping in each quadrant and thereby determines where the spectral weight accumulates. For the type-A interface, the symmetric choices $(\gamma,\gamma')=(0.33,0.33)$ and $(\gamma,\gamma')=(-0.33,-0.33)$ generate the two codimension-two limits: the bulk states are funneled either to the central junction [Fig.~\ref{fig:2D_center_FULL}] or to the four outer corners [Fig.~\ref{fig:2D_corners_FULL}]. When reciprocity is broken only along one direction, the accumulation is reduced to codimension one: for $(\gamma,\gamma')=(0.00,0.33)$ the bulk skin modes pile up on the horizontal interface [Fig.~\ref{fig:2D_midline_FULL}], whereas for $(\gamma,\gamma')=(0.00,-0.33)$ they are redirected to the two outer parallel edges [Fig.~\ref{fig:2D_twoedges_FULL}].

Besides these bulk funneling patterns, the selected cuts also show isolated localized modes superimposed on the skin-state background. The corner and defect states visible in Figs.~\ref{fig:2D_center_FULL}-\ref{fig:2D_twoedges_FULL} remain sharply localized, while the strongly confined states in the weak-coupling cuts [panels (c)] are natural two-dimensional counterparts of the additional interface modes obtained from the interface-coupled solutions in Sec.~\ref{sec3}: they remain detached from the main OBC continuum and become more sharply localized as $t_2$ is reduced. In this sense, the 2D results retain the same basic separation between bulk-like funneling states and interface-originated localized states that emerge from the exact 1D solutions.

\begin{figure*}
  \centering
  \includegraphics[width=0.8\linewidth]{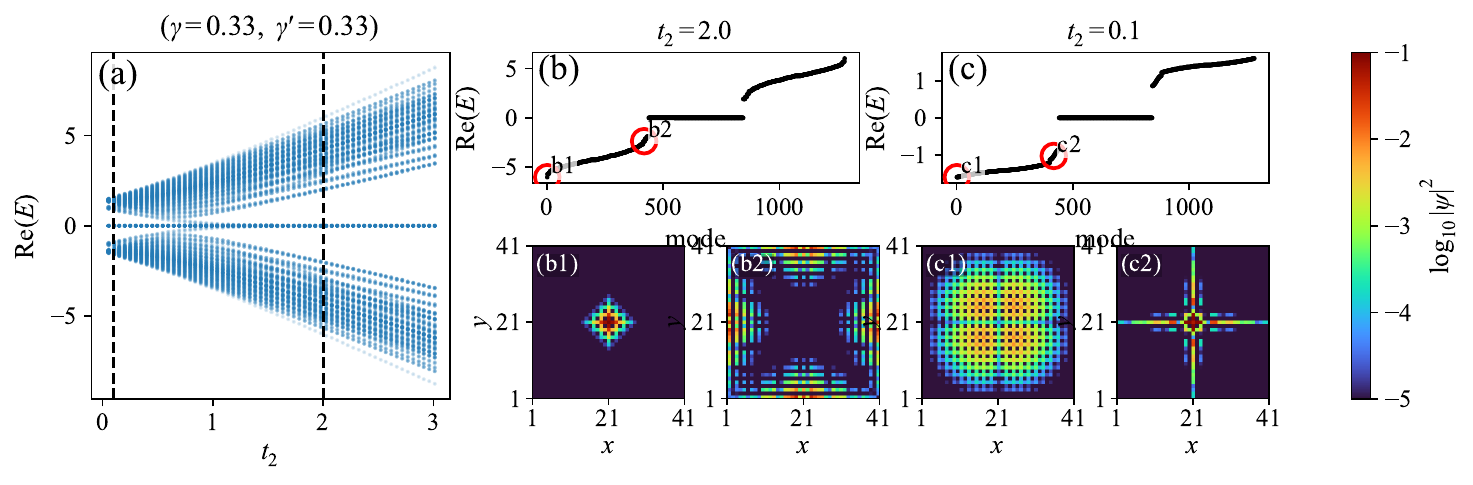}
  \caption{Type-B interface with $(\gamma,\gamma')=(0.33,0.33)$ and $t_1=1$. (a) OBC spectrum versus $t_2$, with representative cuts at $t_2=0.1$ and $t_2=2$. (b) Spectrum at $t_2=2$ and representative right-eigenstate densities showing center-point funneling at the codimension-one junction (b1) and a second localized mode (b2). (c) Spectrum at $t_2=0.1$ together with two localized modes (c1) and (c2).}
  \label{fig:2D_B_center_FULL}
\end{figure*}

\begin{figure*}
  \centering
  \includegraphics[width=0.8\linewidth]{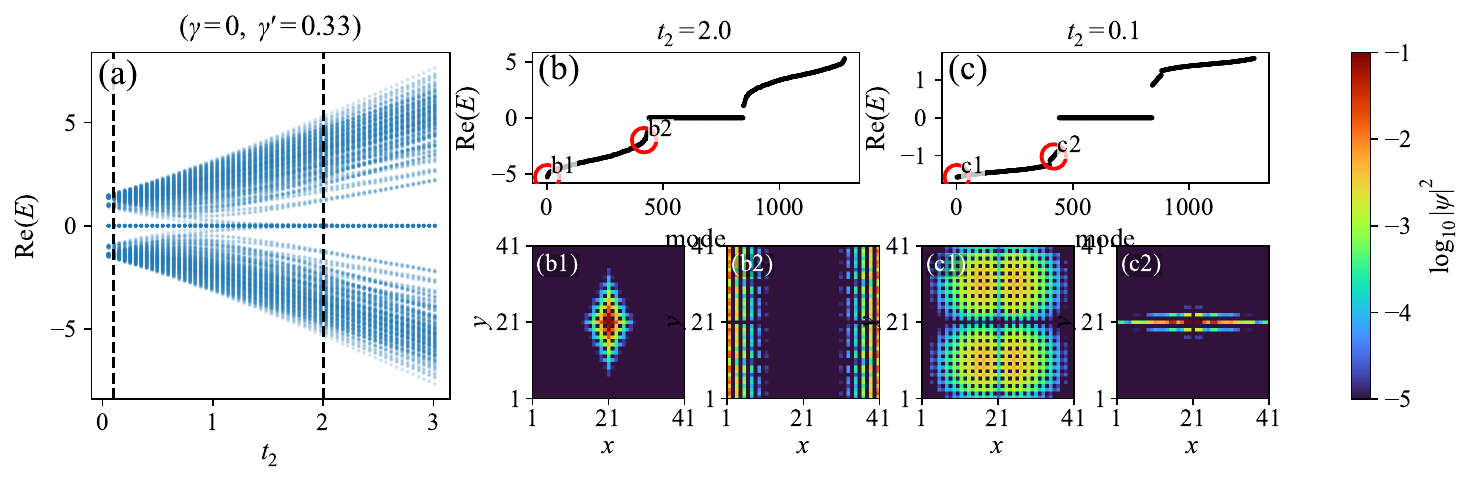}
  \caption{Type-B interface with $(\gamma,\gamma')=(0.00,0.33)$ and $t_1=1$. (a) OBC spectrum versus $t_2$, with representative cuts at $t_2=0.1$ and $t_2=2$. (b) Spectrum at $t_2=2$ and representative right-eigenstate densities showing funneling onto the horizontal codimension-one interface (b1) and a second localized mode (b2). (c) Spectrum at $t_2=0.1$ together with two localized modes (c1) and (c2).}
  \label{fig:2D_B_midline_FULL}
\end{figure*}

We next turn to the type-B interface, which provides a complementary domain-wall geometry to the type-A configuration. Representative type-B cases are shown in Figs.~\ref{fig:2D_B_center_FULL} and \ref{fig:2D_B_midline_FULL}.
Although its detailed connectivity differs, the funneling direction is again controlled by the sign pattern of $(\gamma,\gamma')$. 

For $(\gamma,\gamma')=(0.33,0.33)$, both non-reciprocal biases point inward across the type-B interfaces, driving the bulk states towards their codimension-two intersection at the lattice center, as shown in Fig.~\ref{fig:2D_B_center_FULL}. At $t_2=2$, the representative eigenstates display pronounced accumulation deep in the system interior rather than at the physical boundaries, while additional localized modes remain visible in the spectrum. When the coupling is reduced to $t_2=0.1$, the localization sharpens further, consistent with weaker hybridization away from the interface junction. This confirms that center-point funneling persists in the type-B geometry.

A complementary anisotropic regime appears for $(\gamma,\gamma')=(0.00,0.33)$, shown in Fig.~\ref{fig:2D_B_midline_FULL}. In this case, the system remains Hermitian along $x$ but exhibits NH pumping along $y$, so the accumulation profile is reduced from a codimension-two point to a codimension-one distribution along the horizontal midline. The bulk states are therefore funneled towards the internal interface, while remaining comparatively extended along the orthogonal direction. Together with the center-funneling case above, this shows that the type-B geometry reproduces the same center-to-midline dimensional reduction controlled by the directional pattern of non-reciprocity.

Beyond these bulk skin modes, both Lieb-lattice geometries also host localized states associated with the underlying interface structure. As shown by the representative eigenstate profiles in Figs.~\ref{fig:2D_center_FULL}-\ref{fig:2D_twoedges_FULL} and Figs.~\ref{fig:2D_B_center_FULL}-\ref{fig:2D_B_midline_FULL}, such localized modes coexist with the bulk funneling patterns in both type-A and type-B systems. For example, in the type-A center-funneling regime, a corner-localized mode [Fig.~\ref{fig:2D_center_FULL}(b3)] coexists with the centrally accumulated bulk states. At smaller $t_2$, additional strongly localized states become clearly separated from the bulk-funneling background in both interface geometries.

\section{Conclusion}
We have presented a systematic analytical study of domain walls in the nonreciprocal SSH chain. By constructing inversion-symmetric interface geometries (type A and type B) from two segments with inverted coupling ratios, we derived exact quantization conditions and eigenstates under open boundaries using two complementary ansatzes. Our results clarify how the non-Hermitian skin effect reshapes interface physics: beyond the skin-mode continuum and the chiral-symmetry-protected zero modes inherited from the SSH building blocks, domain walls can host additional finite-energy states that remain isolated from the bulk OBC spectrum and become extended only at fine-tuned parameter values.

More broadly, our work connects to the ongoing effort to restore and generalize the bulk-boundary correspondence in non-Hermitian systems~\cite{kawabata2019prx,bergholtz2021exceptional}. In particular, domain-wall problems provide a natural setting to test non-Bloch formulations~\cite{yao2018edge,yokomizo2019non} and to define interface-sensitive topological diagnostics that go beyond edge spectra~\cite{okuma2020topological,zhang2020correspondence}. A promising direction is to identify an invariant that predicts, in a unified way, (i) the existence of the additional finite-energy interface states reported here and (ii) their complementary appearance in type-A versus type-B geometries~\cite{bergholtz2021exceptional}. It would also be interesting to formulate these interface modes in terms of transfer matrices~\cite{kunst2019non} and exceptional-point physics~\cite{bergholtz2021exceptional}, and to study how their localization and spectral isolation evolve in the presence of disorder, nonlinearity, interactions, or quasiperiodic modulations~\cite{ashida2020non,Wu2026Nonlinear}.

We further extended the interface construction to a two-dimensional Lieb lattice with independent nonreciprocities along the two directions. The resulting codimension-one and codimension-two interfaces enable controllable ``funneling'' of bulk states towards midlines, corners, or a central junction, coexisting with symmetry-protected boundary and defect modes~\cite{weidemann2020topological,song2020two,lee2019hybrid}. Moving forward, higher-dimensional generalizations may reveal a hierarchy of ``higher-order'' skin and interface phenomena (e.g., bulk-to-hinge and bulk-to-corner funneling)~\cite{lee2019hybrid,zhang2021observation, Phan2026Corner} that can be tuned by interface geometry, anisotropic nonreciprocity, and lattice symmetries~\cite{kawabata2019prx, bergholtz2021exceptional}. From a dynamical perspective, it will be valuable to quantify how these steady-state localization patterns manifest in wave-packet evolution, driven-dissipative settings, and response measurements~\cite{xiao2020non,liang2022dynamic}.

Finally, our exact solutions provide a transparent route towards engineering non-Hermitian localization landscapes and may be useful for experimental platforms ranging from photonics~\cite{parto2021non} and topolectrical circuits~\cite{helbig2020generalized,imhof2018topolectrical} to cold atoms~\cite{liang2022dynamic} and quantum walks~\cite{weidemann2020topological}. We expect that controlled domain-wall engineering will continue to be a versatile tool for manipulating non-Hermitian transport~\cite{ashida2020non} and for exploring interface topological physics beyond conventional Hermitian paradigms~\cite{bergholtz2021exceptional}.

\begin{acknowledgments}
T.W. and F.K.K. acknowledge funding from the Max Planck Society Lise Meitner Excellence Program~\mbox{2.0}.
\end{acknowledgments}

\section*{Data Availability}
The data that support the findings of this article are not publicly available. The numerical results were generated using custom Python, MATLAB, and Mathematica code, which is available from the corresponding author upon reasonable request.

\appendix
\section{Biorthogonal Polarization for skin modes}
\label{appendix}

\begin{figure*}
    \centering
    \includegraphics[width=0.7\linewidth]{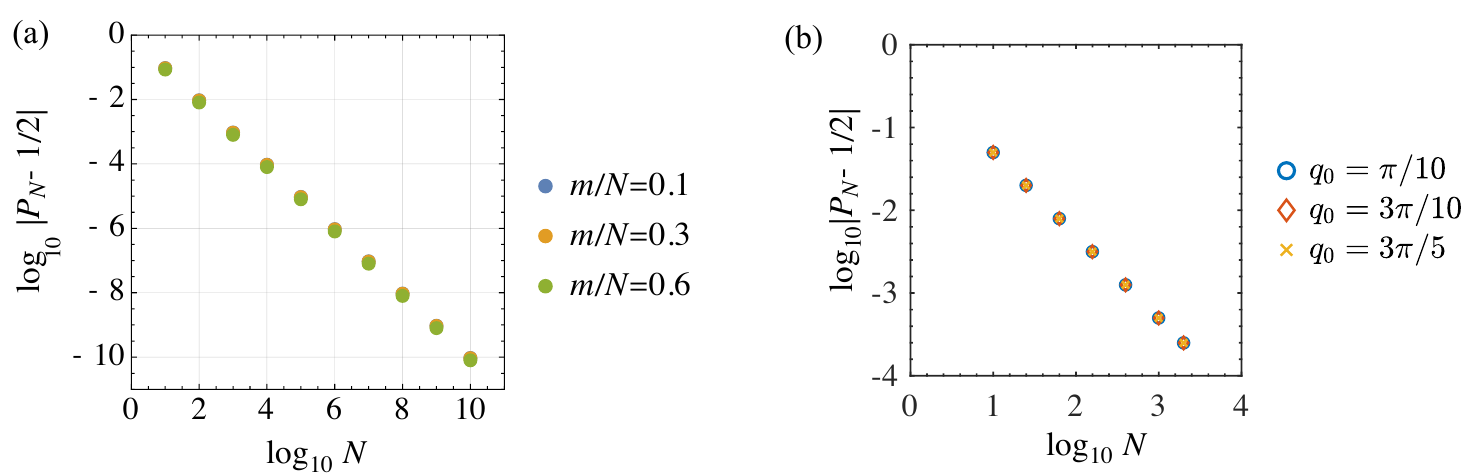}
    \caption{Biorthogonal polarization $P_N$ as a function of system size $N$ for the (a) broken- and (b) unbroken-unit-cell case. We fix the system parameters at $t_1=0.6$, $t_2=1$, and $\gamma=0.2$. (a) The blue, orange, and greens dots are for eigenstates with different $q_m$, where $m$ scales with the system size $N$ to keep the value of $q_m$ almost fixed. (b) The circle, diamonds and crosses are for eigenstates with $q$ closest to different initialization $q_0$.}
    \label{fig:appendix_PN}
\end{figure*}

\begin{figure*}
    \centering
    \includegraphics[width=0.9\linewidth]{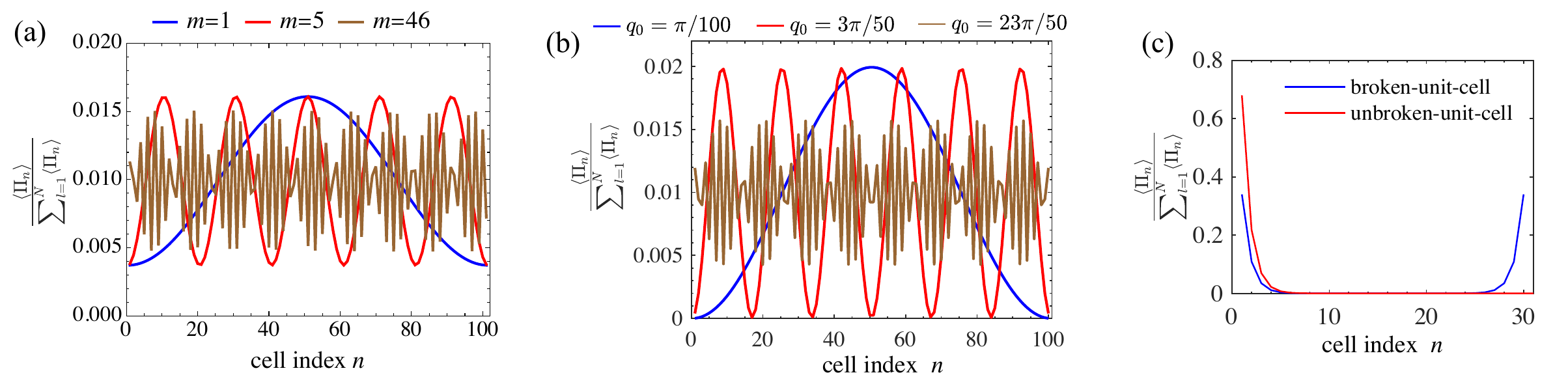}
    \caption{Biorthonormal expectation value of the projector for different eigestates of the NH SSH model for the (a) broken-, (b) unbroken-unit-cell case, and (c) the zero modes in both cases in blue and red, respectively. We fix the system size $N=100$, and parameters $t_1=0.6$, $t_2=1$, and $\gamma=0.2$. The blue, red and brown lines are for eigenstates (a) with different $q_m$ and (b) with $q$ closest to different initialization $q_0$.}
    \label{fig:appendix_projector}
\end{figure*}

In this appendix we show that the biorthogonal polarization $P$ defined in Eq.~\eqref{eq_biortho_polarization} for the skin states of the NH SSH model always takes the value $1/2$. The right eigenstate $|\psi_{R,\pm}(q)\rangle$ with a generalized wave vector $q$ that satisfies the OBC constraints is given in Eqs.~\eqref{eq_wf_total} and \eqref{eq_wf_component}. Here the subscript $\pm$ labels the two OBC bands in Eq.~\eqref{eq_Eq}. By making the substitution $\gamma\to-\gamma$ one can obtain the corresponding left eigenstate 
\begin{align}
    \bar{\Psi}_{L,\pm,n}(q) &= \langle\Psi_{L,\pm}(q)|e_n\rangle \nonumber\\
    &= r^{-n} [e^{iqn} \bar{\psi}_{L,\pm,n}(q)- e^{-iqn}\bar{\psi}_{L,\pm,n}(-q)],
\end{align}
where
\begin{align}
    \bar{\psi}_{L,\pm}(q) = \left[
    t_1 - \gamma + rt_2e^{-iq},\ E_{\pm}^{\rm OBC}(q) \right].
\end{align}
With the unbiorthonormalized left and right states, one can compute the expectation value of the projection operator $\Pi_n = c^{\dagger}_{A,n} \ket{0} \bra{0}c_{A,n}^{\vphantom{\dagger}} + c^{\dagger}_{B,n} \ket{0} \bra{0}c_{B,n}^{\vphantom{\dagger}}$ as
\begin{align}
    \langle\Pi_n\rangle &=\langle \Psi_{L,\pm}(q)|\Pi_n|\Psi_{R,\pm}(q)\rangle \nonumber\\
    &=2t_2 \sqrt{t_1^2-\gamma^2}\left[\sin(q n)\sin(q(n-1))+\cos q\right]\nonumber\\
    &+ 2(t_1^2-\gamma^2)\sin^2(q n)  \nonumber\\
    &+ t_2^2[\sin^2(q n)+\sin^2(q (n-1))].
\end{align}
To study the asymptotic behavior of $P$, we define the biorthogonal polarization of an eigenstate in a finite chain with $N$ unit cells as
\begin{equation}
    P_N = 1 - \frac{\langle \Psi_L| \sum_{n=1}^N n\Pi_n|\Psi_R\rangle}{N \langle\Psi_L|\Psi_R\rangle},
\end{equation}
where we added $\langle\Psi_L|\Psi_R\rangle$ in the denominator to biorthonormalize the eigenstates.
It can be rewritten as
\begin{equation}
    P_N = 1-\frac{1}{N}\frac{\sum_{n=1}^N n\langle\Pi_n\rangle}{\sum_{n=1}^N \langle\Pi_n\rangle}.
\end{equation}

First we focus on the broken-unit-cell case, where the generalized wave vector $q$ has the explicit expression
\begin{equation}
    q_m = \frac{m \pi}{N+1},\quad m=1,2,\cdots ,N.
\end{equation}
Here $N$ is the number of intact unit cells in the chain. In Fig. \ref{fig:appendix_PN}(a) we can see that $P_N$ for different values of $q_m$ converge to $1/2$ unanimously with a power-law. The slope of the data points in the log-log plot is almost 1, indicating $P_N-1/2\propto N^{-1}$. We further show in Fig.~\ref{fig:appendix_projector}(a) the distribution of $\langle \Pi_n \rangle$ over the lattice sites in a chain of size $N=100$. Similar to the wave functions of bulk states in Hermitian systems, the projectors of the skin states in the biorthonormal basis are extended over the entire lattice and a larger index $m$ leads to faster sinusoidal modulations. The extended nature of $\langle \Pi_n \rangle$ for the skin modes naturally leads to the result that $P=\lim_{N\to\infty}P_N=1/2$ as well as its power-law convergence. 

Next we turn to the unbroken-unit-cell case, where the wave numbers $q$ have no explicit expression but are solutions of the quantization condition in Eq.~\eqref{eq_q_quantization} and are complex numbers in principal. Therefore, we turn to numerical calculations instead. For a given eigenstate of index $1 \le m \le N$, we solve for $q_m$ numerically in the complex domain with an initial guess $q_0 = m\pi/(N + 1)$, and then compute the projectors and $P_N$ using the solution. Again $P_N$ features the same power-law convergence towards $1/2$ as shown in Fig. \ref{fig:appendix_PN}(b), and the biorthogonal projections of the skin modes are extended over the lattice as shown in Fig. \ref{fig:appendix_projector}(b). In (c) we further show the biorthogonal projectors for the zero-energy mode in both the broken- and unbroken-unit-cell cases in blue and red, respectively, in a lattice with $30$ unit cells. While the zero mode in the broken-unit-cell case is localized at a certain end, the topological zero mode in the unbroken-unit-cell case is localized on both ends of the open chain, leading to $P=1/2$ as in the case for skin modes. The results presented in this appendix are consistent with Eq.~\eqref{eq:top_phase_ssh} and related discussions in Sec. \ref{sec2c}, and explain the behaviors of the biorthogonal polarization $P$ for different eigenstates of the NH SSH model shown in Fig.~\eqref{fig:biortho_polarization_single_chain}.

\bibliography{reference.bib}
\end{document}